\documentclass{aa}
\usepackage{graphicx}
\usepackage{txfonts}
\usepackage[figuresright]{rotating}
\usepackage{natbib}
\bibpunct{(}{)}{;}{a}{}{,}

\def\ngc{\object{NGC\,6231}}

\def\cpd{\object{CPD\,$-$\,41\degr\,7742}}
\def\hda{\object{HD\,152248}}

\def\kms{km\ s$^{-1}$}
\def\cnts{cnt\ s$^{-1}$}
\def\ergs{erg\ s$^{-1}$}
\def\ergscm{erg\ cm$^{-2}$\ s$^{-1}$}
\def\rsol{R$_{\odot}$}
\def\msol{M$_{\odot}$}
\def\l{$\lambda$}

\def\lxlbol{$L_\mathrm{X} / L_\mathrm{bol}$}

\def\hea{\ion{He}{i}\ }
\def\heb{\ion{He}{ii}\ }

\def\xmm{{\sc XMM}\emph{-Newton}}

\def\mos{{\sc MOS}}
\def\pn{pn}

\def\epicmos{{\sc EPIC MOS}}
\def\epicpn{{\sc EPIC} pn}

\def\xspec{{\sc xspec}}

\def\sas{{\sc sas}}

\defcitealias{SHRG03}{Paper~I}

\begin{document}
   \title{The massive binary CPD\,$-$\,41\degr\,7742 :  }

   \subtitle{II. Optical light curve and X-ray observations \thanks{Based on observations collected at the European Southern Observatory (La Silla, Chile) and with \xmm, an ESA Science Mission with instruments and contributions  directly funded by ESA Member States and the USA (NASA).}}

   \author{ 
H. Sana\inst{1}\fnmsep\thanks{Research Fellow FNRS (Belgium)}
	\and
E. Antokhina\inst{2}
	\and
P. Royer\inst{3}
        \and
J. Manfroid\inst{1}\fnmsep\thanks{Research Director FNRS (Belgium)}
	\and
E. Gosset\inst{1}\fnmsep\thanks{Research Associate FNRS (Belgium)}
	\and
G. Rauw\inst{1}\fnmsep$^{\dagger}$ 
        \and
J.-M. Vreux\inst{1}
          }

   \offprints{H. Sana}

   \institute{Institut d'Astrophysique et de G\'eophysique, 
              Universit\'e de Li\`ege, All\'ee du 6 Ao\^ut 17, B\^at. B5c,
              B-4000 Li\`ege, Belgium\\
              \email{sana@astro.ulg.ac.be,
		     manfroid@astro.ulg.ac.be,  gosset@astro.ulg.ac.be, \\rauw@astro.ulg.ac.be,
		     vreux@astro.ulg.ac.be}
         \and
	      Sternberg Astronomical Institute, Moscow State University, \\
	      Universitetskii pr., 13, 119899 Moscow, Russia 
              \email{elant@sai.msu.ru}
	 \and
              Instituut voor Sterrenkunde, Katholieke Universiteit Leuven,
	      Celestijnenlaan 200 B, B-3001 Leuven, Belgium\\
              \email{pierre@ster.kuleuven.ac.be} 
             }

   \date{Received September 15, 2000; accepted September 15, 2000}

   \abstract{
In the first paper of this series, we presented a detailed high-resolution spectroscopic study of \cpd, deriving for the first time an orbital solution for both components of the system. In this second paper, we focus on the analysis of the optical light curve and on recent \xmm\ X-ray observations. In the optical, the  system presents two eclipses, yielding an inclination $i\sim77\degr$.
Combining the constraints from the photometry with the results of our previous work, we derive the absolute parameters of the system. We confirm that the two components of \cpd\ are main sequence stars (O9~V + B1-1.5~V) with masses ($M_1\sim18$~\msol\ and $M_2\sim10$~\msol) and respective radii  ($R_1\sim7.5$~\rsol\ and $R_2\sim5.4$~\rsol) close to the typical values expected for such stars.

We also report an unprecedented set of X-ray observations that almost uniformly cover the 2.44-day orbital cycle. The X-ray emission from \cpd\ is well described by a two-temperature thermal plasma model with energies close to 0.6 and 1.0\,keV, thus slightly harder than typical early-type emission. The X-ray light curve shows clear signs of variability. The emission level is higher when the primary is in front of the secondary. During the high emission state, the system shows a drop of its X-ray emission that almost exactly matches the optical eclipse. We interpret the main features of the X-ray light curve as the signature   of a wind-photosphere interaction, in which the overwhelming primary O9 star wind crashes into the secondary surface. Alternatively the light curve could result from a wind-wind interaction zone located near the secondary star surface. As a support to our interpretation, we provide a phenomenological geometric model that qualitatively reproduces the observed modulations of the X-ray emission. 
   }

\titlerunning{The Massive Binary CPD\,$-$\,41\degr\,7742. II}
\authorrunning{H. Sana et al.}

   \maketitle
%

\begin{table}
\centering
\caption{Orbital and physical parameters of \cpd\ as derived from the \hea 
lines orbital solution presented in \citetalias{SHRG03}. The usual notations 
have been adopted. $T_0$ is the time of periastron passage and is adopted as 
phase $\psi=0.0$ . The column to the right provides the revised estimate  of 
the errors, obtained with Monte-Carlo simulation techniques (see Sect. 
\ref{sect: discuss_photo}). }
\label{tab: orbit}
\begin{tabular}{l c c c | c}
\hline
\hline
$P_\mathrm{orb}$ (days)& 2.44070 & $\pm$ & 0.00050 & \\
$e$                   &    0.027 & $\pm$ & 0.006   & 0.008 \\
$\omega$ (\degr)      &    149   & $\pm$ & 10      & 17  \\    
$T_0$(HJD             & 2400.284 & $\pm$ & 0.067   & 0.113 \\
\hspace*{2mm} $-$2\,450\,000) &  &                 & \\
$\gamma_1$ (\kms)     &  $-15.3$ & $\pm$ & 0.5     & 1.2 \\
$K_1$ (\kms)          &    167.1 & $\pm$ & 0.9     & 1.4  \\
$a_1\sin i$ (\rsol)   &    8.05  & $\pm$ & 0.05    & 0.07 \\ 
$\gamma_2$ (\kms)     &  $-26.3$ & $\pm$ & 0.7     & 2.4 \\
$K_2$ (\kms)          &    301.3 & $\pm$ & 1.8     & 3.0 \\
$a_2\sin i$ (\rsol)   &    14.52 & $\pm$ & 0.09    & 0.14 \\
                      &          &       &         &      \\
$q\ (=M_1/M_2)$       &    1.803 & $\pm$ & 0.015   & 0.023 \\
$M_1\sin^3 i$ (\msol) &    16.69 & $\pm$ & 0.25    & 0.39 \\
$M_2\sin^3 i$ (\msol) &    9.25  & $\pm$ & 0.12    & 0.18 \\		
\hline 
\end{tabular}
\end{table}

\section{Introduction \label{sect: intro}}

In the quest for accurate measurements of fundamental stellar parameters, 
eclipsing spectroscopic binaries are unique physical laboratories all over the Hertzsprung-Russell diagram. Combined spectroscopic and photometric studies provide a direct  determination of the masses and sizes of their stellar components. This is of a particular interest in the upper left part of the diagram. Although few in number, massive early-type stars  have a large influence on their surroundings through their mechanical and radiative energy input. A detailed knowledge of both their evolution and wind properties is thus crucial in many different contexts. For example, these objects  seem to play a key role in the formation of the less massive stars in starburst regions or within the core of OB associations. However, our understanding of massive stars is clearly still fragmentary. Only a few tens of objects have their orbital and physical parameters determined with a reasonable accuracy \citep{Gie03}. The problem of their exact formation mechanism  is largely unsolved \citep{Zin03} and, from the theoretical point of view, their physical parameters (effective temperatures, radii, masses, ...) significantly differ from one study to another \citep{HM84, HP89, VGS96}.  The observational masses deduced from atmosphere models  are systematically lower than the predicted masses using evolutionary models (the so-called {\it mass discrepancy} problem, \citealt{HKV92, He03}). Fortunately, recent works \citep{CHE02, HPN02, BG02, MSH02} using line-blanketed atmosphere models and accounting both for the spherical stellar atmosphere and for the stellar winds yielded new effective temperature scales for early-type stars and, simultaneously, led to a better agreement between the spectroscopic and evolutionary masses. In this context, the accurate determination of the massive star fundamental parameters, over the whole spectral type and luminosity class range covered by these objects, provides thus the basic material to strengthen our understanding of this particularly important stellar population. 

The early-type binary systems are also crucial for the mapping of X-ray emitting plasmas. So far, the most reliable way to constrain the geometry of the hot plasma around stars of various spectral types is through the study of the temporal changes of the X-ray fluxes of eclipsing binaries or rotating stars; the latter only in cases of non-uniform surface distributions of X-ray plasma. A good time coverage of the orbital or rotation cycle is of course critical to provide as complete a description as possible. While late-type stars often experience flaring activities which may considerably complicate the task of mapping their coronae, the situation should, in principle, be much easier in early-type stars. In fact, single early-type objects usually do not display a strong X-ray variability \citep{BS94}. In early-type binaries, a significant fraction of the X-ray emission may however arise in a  wind interaction zone. The orbital modulation of their X-ray flux is thus quite common, either because of the changing opacity along the line of sight towards the shock region, or as a consequence of the changing properties of the wind interaction zone in an eccentric binary.

In this context, we have undertaken a detailed study of \cpd, a double line spectroscopic binary located in the core of the young open cluster \ngc. In \citet[ Paper I hereafter]{SHRG03}, we presented a first accurate orbital solution for the two components of the system. We derived a short period $P=2.44070$ days and a slight but definite eccentricity $e=0.027$. Based on spectroscopic criteria, we proposed a spectral type and a luminosity class of O9~III + B1~III for the two components of the system. However we outlined the strong ambiguity concerning the quoted luminosity classification. Indeed the luminosities and radii inferred from the membership in \ngc\  rather indicate a class V or IV for both components. The analysis of the light curve of the system will allow us to elucidate this question. 

 We refer to \citetalias{SHRG03} for a review of the previous works on the object. In \citetalias{SHRG03}, we dit not mention the work of \citet{BL95} in which the authors present a first light curve of \cpd, showing a clearly-marked eclipse. We also refer to \citetalias{SHRG03} for details on the spectroscopic analysis of the system. Table \ref{tab: orbit} summarizes the computed orbital solution and the constraints obtained on its physical parameters. This second paper will complete our current view of the system by providing the analysis of the  photometric light curve and of \xmm\  X-ray  observations. It is organised as follows. After a description of the optical and X-ray data sets and data handling (Sect. \ref{sect: obs}), we present the analysis of the system light curve (Sect. \ref{sect: lc}). In Sect. \ref{sect: discuss_photo}, we combine the newly obtained information  with results from \citetalias{SHRG03} and we derive the absolute parameters of the system. The X-ray properties of \cpd\ are described in Sect. \ref{sect: X}.  In Sect. \ref{sect: Xmodel}, we investigate the wind properties and we propose to interpret the X-ray light curve as the signature of a wind interaction. We also present a simple phenomenological model that reproduces reasonably well  the observed modulations.  Final considerations and  conclusions of this work are summarised in Sect. \ref{sect: ccl}.

\begin{table*}
\centering
\caption{Journal of the photometric observations of \cpd\ in the two filters \l6051 and \l4686 (see text). Odd columns give the heliocentric Julian dates (in format HJD $-$ 2\,450\,000). Even columns provide the observed magnitudes in the selected filter.}
\label{tab: photo}
\begin{tabular}{c c c c c c |  c c c c c c}
\hline
\hline
\multicolumn{6}{c|}{\l6051} & \multicolumn{6}{c}{\l4686} \\
HJD & mag & HJD & mag & HJD & mag & HJD & mag & HJD & mag & HJD & mag \\
\hline
 530.8192 & 8.103 & 540.8836 & 8.103 & 549.9225 & 8.417 &    534.8722 & 8.540 &  545.8333 & 8.533 & 554.8312 & 8.777 \\
 530.8203 & 8.098 & 540.9142 & 8.088 & 550.7756 & 8.129 &    534.8968 & 8.533 &  545.8758 & 8.525 & 554.8680 & 8.678 \\
 531.8341 & 8.130 & 540.9223 & 8.098 & 550.8198 & 8.132 &    535.7976 & 8.520 &  545.9226 & 8.546 & 554.9149 & 8.594 \\
 531.8354 & 8.123 & 540.9286 & 8.101 & 550.8595 & 8.129 &    535.8262 & 8.526 &  546.7415 & 8.521 & 555.7662 & 8.569 \\
 533.8018 & 8.123 & 541.7697 & 8.106 & 550.8872 & 8.125 &    535.8670 & 8.534 &  546.7979 & 8.516 & 555.8110 & 8.574 \\
 533.8033 & 8.127 & 541.7939 & 8.105 & 550.9163 & 8.139 &    535.9103 & 8.530 &  546.8556 & 8.522 & 555.8573 & 8.643 \\
 533.8041 & 8.126 & 541.8431 & 8.108 & 550.9216 & 8.148 &    537.7754 & 8.690 &  546.9197 & 8.527 & 555.9148 & 8.726 \\
 533.8291 & 8.138 & 541.8689 & 8.091 & 551.7497 & 8.105 &    537.8168 & 8.607 &  547.7870 & 8.541 & 555.9303 & 8.756 \\
 533.8299 & 8.147 & 541.8974 & 8.112 & 551.7937 & 8.115 &    537.8576 & 8.568 &  547.8468 & 8.542 & 556.7881 & 8.535 \\
 533.8307 & 8.146 & 542.7511 & 8.127 & 551.8060 & 8.124 &    537.8590 & 8.571 &  547.8905 & 8.517 & 556.8690 & 8.558 \\
 533.8571 & 8.161 & 542.7896 & 8.128 & 551.8631 & 8.124 &    537.9009 & 8.552 &  547.9242 & 8.510 & 557.7451 & 8.517 \\
 533.8579 & 8.160 & 542.8192 & 8.141 & 551.9181 & 8.100 &    538.7595 & 8.614 &  548.7645 & 8.692 & 557.7899 & 8.525 \\
 533.8587 & 8.151 & 542.8482 & 8.118 & 552.7605 & 8.113 &    538.8022 & 8.674 &  548.8095 & 8.614 & 557.8334 & 8.525 \\
 533.8823 & 8.198 & 542.8906 & 8.126 & 552.8081 & 8.117 &    538.8615 & 8.755 &  548.8509 & 8.565 & 557.8599 & 8.509 \\
 533.8831 & 8.188 & 543.7525 & 8.342 & 552.8566 & 8.105 &    538.9045 & 8.772 &  548.8781 & 8.544 & 557.9078 & 8.506 \\
 533.8839 & 8.194 & 543.8084 & 8.356 & 552.8789 & 8.093 &    539.7730 & 8.560 &  548.9168 & 8.538 & 557.9249 & 8.511 \\
 533.9053 & 8.234 & 543.8701 & 8.292 & 552.8798 & 8.093 &    539.8246 & 8.552 &  548.9212 & 8.538 & 558.6961 & 8.545 \\
 533.9063 & 8.236 & 543.8955 & 8.242 & 552.9189 & 8.090 &    539.8655 & 8.560 &  549.7466 & 8.710 & 558.7425 & 8.543 \\
 533.9270 & 8.273 & 543.9226 & 8.199 & 553.7507 & 8.118 &    539.8983 & 8.569 &  549.8067 & 8.858 & 558.7897 & 8.521 \\
 533.9278 & 8.263 & 544.7848 & 8.159 & 553.7947 & 8.124 &    540.7890 & 8.514 &  549.8637 & 8.902 & 558.8082 & 8.543 \\
 533.9286 & 8.267 & 544.8474 & 8.255 & 553.8328 & 8.116 &    540.8306 & 8.516 &  549.9214 & 8.824 &	  &	   \\
 533.9294 & 8.266 & 544.8936 & 8.374 & 553.8736 & 8.114 &    540.8815 & 8.523 &  550.7748 & 8.534 &	  &	   \\
 533.9302 & 8.274 & 544.9210 & 8.442 & 553.9195 & 8.111 &    540.9135 & 8.531 &  550.8190 & 8.531 &	  &	   \\
 534.7974 & 8.123 & 544.9281 & 8.457 & 554.7069 & 8.470 &    540.9217 & 8.547 &  550.8586 & 8.537 &	  &	   \\
 534.8227 & 8.129 & 545.7752 & 8.112 & 554.7540 & 8.485 &    540.9280 & 8.542 &  550.8864 & 8.524 &	  &	   \\
 534.8456 & 8.130 & 545.8341 & 8.119 & 554.7937 & 8.430 &    541.7689 & 8.523 &  550.9155 & 8.546 &	  &	   \\
 534.8730 & 8.142 & 545.8766 & 8.123 & 554.8319 & 8.343 &    541.7931 & 8.534 &  550.9208 & 8.561 &	  &	   \\
 534.8976 & 8.132 & 545.9234 & 8.128 & 554.8686 & 8.253 &    541.8423 & 8.532 &  551.7489 & 8.528 &	  &	   \\
 535.7984 & 8.118 & 546.7423 & 8.100 & 554.9156 & 8.168 &    541.8681 & 8.523 &  551.7930 & 8.535 &	  &	   \\
 535.8270 & 8.118 & 546.7987 & 8.097 & 555.7669 & 8.148 &    541.8966 & 8.538 &  551.8052 & 8.528 &	  &	   \\
 535.8678 & 8.124 & 546.8564 & 8.095 & 555.8117 & 8.168 &    542.7503 & 8.546 &  551.8623 & 8.535 &	  &	   \\
 535.9111 & 8.110 & 546.9205 & 8.109 & 555.8581 & 8.228 &    542.7888 & 8.542 &  551.9173 & 8.519 &	  &	   \\
 537.7762 & 8.275 & 547.7881 & 8.108 & 555.9156 & 8.310 &    542.8183 & 8.569 &  552.7569 & 8.542 &	  &	   \\
 537.8176 & 8.195 & 547.8476 & 8.109 & 555.9311 & 8.342 &    542.8474 & 8.535 &  552.7614 & 8.540 &	  &	   \\
 537.8601 & 8.157 & 547.8913 & 8.086 & 556.7887 & 8.135 &    542.8898 & 8.546 &  552.8074 & 8.528 &	  &	   \\
 537.9021 & 8.145 & 547.9250 & 8.092 & 556.8698 & 8.138 &    543.7516 & 8.769 &  552.8551 & 8.530 &	  &	   \\
 538.7603 & 8.186 & 548.7674 & 8.255 & 557.7458 & 8.106 &    543.8076 & 8.765 &  552.8782 & 8.521 &	  &	   \\
 538.8030 & 8.258 & 548.8105 & 8.184 & 557.7907 & 8.095 &    543.8693 & 8.701 &  552.9174 & 8.522 &	  &	   \\
 538.8623 & 8.328 & 548.8519 & 8.140 & 557.8342 & 8.087 &    543.8947 & 8.657 &  553.7500 & 8.557 &	  &	   \\
 538.9053 & 8.342 & 548.8789 & 8.115 & 557.8607 & 8.082 &    543.9218 & 8.623 &  553.7941 & 8.556 &	  &	   \\
 539.7738 & 8.137 & 548.9179 & 8.115 & 557.9094 & 8.079 &    544.7840 & 8.560 &  553.8321 & 8.540 &	  &	   \\
 539.8254 & 8.132 & 548.9200 & 8.119 & 557.9257 & 8.087 &    544.8465 & 8.672 &  553.8729 & 8.554 &	  &	   \\
 539.8663 & 8.140 & 548.9223 & 8.113 & 558.6969 & 8.127 &    544.8928 & 8.784 &  553.9188 & 8.550 &	  &	   \\
 539.8991 & 8.142 & 549.7474 & 8.302 & 558.7433 & 8.103 &    544.9202 & 8.858 &  554.7062 & 8.887 &	  &	   \\
 540.7898 & 8.096 & 549.8075 & 8.449 & 558.7910 & 8.099 &    544.9272 & 8.863 &  554.7533 & 8.914 &	  &	   \\
 540.8314 & 8.101 & 549.8645 & 8.490 & 558.8091 & 8.113 &    545.7744 & 8.530 &  554.7928 & 8.840 &	  &	   \\
\hline					    
\end{tabular}
\end{table*}


\begin{figure}[htb]
\centering
\includegraphics[width=8.5cm]{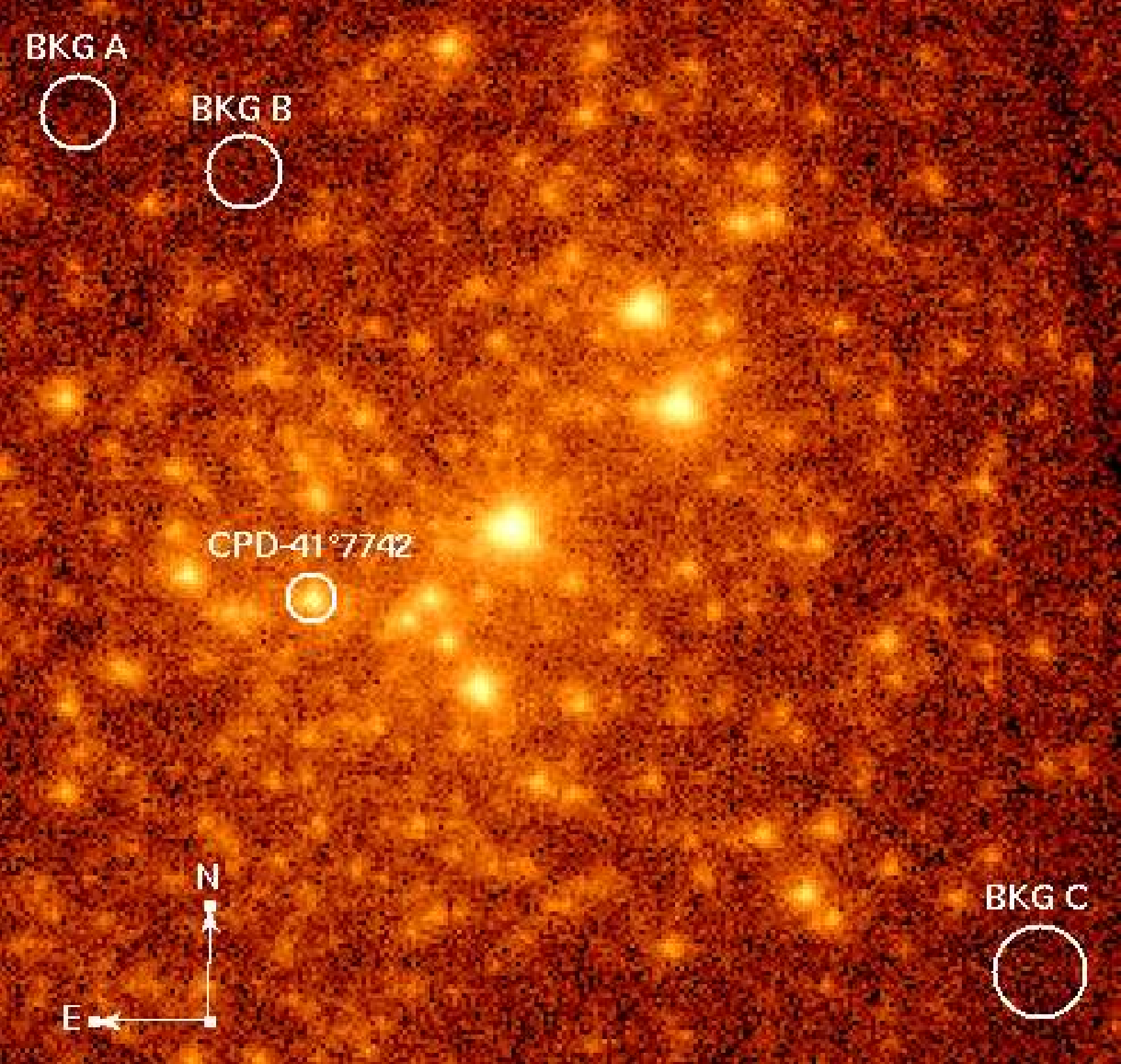}
\caption{Broad-band [0.5 - 10.0\,keV] image of the \ngc\ core. This \epicmos1+\mos2 image combines the two instruments and  the six pointings of the campaign for a cumulated effective exposure time of 351.5\,ks.  The source and background extraction regions are shown.}
\label{fig: fov}
\end{figure}

\section{Observations and data reduction \label{sect: obs}}

\subsection{Photometry \label{ssect: obs_photo}}
Between 1997 March 22 and April 19, we observed
the core of the open cluster \ngc\
with the 0.6-m Bochum telescope at La Silla, Chile.
The Cassegrain focus of the telescope was equipped
with a direct camera and a Thomson 7882 charge-coupled
device (CCD) detector (384 $\times$ 576 pixels) subtending
a full field of view of 3.2 by 4.8 arcmin.
The photometric observations have been performed through two
narrow band filters: one called $\lambda$4686 addressing
the region of the \heb line usually
present in massive stars
(centre: 4684 \AA , FWHM: 30 \AA ) and
another one labelled $\lambda$6051 addressing a region
of the continuum free from strong lines
(centre: 6051 \AA , FWHM: 28 \AA ).
More information on these filters can be found in 
\citet{RVM98}.
The typical exposure times were 60s for both filters.
Some 112 (resp.\ 138) useful frames were obtained with the
$\lambda$4686 (resp.\ $\lambda$6051) filter.
Flat field calibrations were obtained daily on the floodlit dome. 
No twilight flat could be
acquired due to the narrowness of the filters.
Several biases were cautiously acquired at various times during the 
different nights.
The frames were debiased using a master zero frame and,
in the absence of overscan, a level value interpolated
between the various bias frames taken during the same night.
The optical elements close to the CCD proved to be
frequently contaminated by dust. Hence, the pixel-to-pixel
(high spatial frequency)
part of the flat-field calibration had to be carefully
extracted from the calibration frames obtained daily.
The large scale component of the dome flat fields
varied slightly from day to day. This was found to be 
due to minor changes in the instrumental
setup. Night sky superflats proved to be more stable,
but yielded a strong systematic vignetting as shown
by \citet{MRR01}.
Consequently, the illumination correction
was entirely obtained from the `photometric superflats'
based on stellar measurements \citep[see e.g.][]{Man95}.
All reductions were carried out with the
National Optical Astronomy Observatories (NOAO) {\sc{iraf}}
package.
The debiased, flat-fielded frames were analyzed
with the {\sc{daophot}} software \citep{Ste87}, using 
aperture radii between 2 and 5.5 arcsec.
`Absolute' photometry was derived from the large aperture
data, using a multi-night, multi-filter algorithm and
a few standard stars \citep{Man93}.
This procedure yielded additional reference stars for each field.
These secondary standards together with all non-variable
stars were used to fix, through a global minimization procedure,
the zero points for the individual frames
and for each aperture radius, thus performing some kind of
global differential photometry.

Comparing the photometry performed through the different
apertures, we noted that a faint companion visible
at 3-4\arcsec\ to the W-SW of \cpd\ has actually
no influence on the differential photometry.
The final magnitudes are given in Table \ref{tab: photo} and correspond
to a 2\farcs 5 radius aperture. The expected error on a star
of similar brightness as \cpd\ corresponds to
$\sigma$~=~0.007 mag in differential photometry.

\subsection{X-ray observation \label{ssect: obs_xray}}
\cpd\ was observed with \xmm\ \citep{Jansen01_xmm} during the six pointings of the campaign towards \ngc\ \citep{SSG04, SGR05} performed within the guaranteed time programme of the Optical Monitor consortium. The \mos\ cameras \citep{Turner01_mos} were operated in the full frame mode and using the thick filter to avoid contamination by UV/optical light. No \epicpn\ data were collected for \cpd\ since the star fell on a gap of the \pn\ detector. Due to the brightness of the objects in the field of view (FOV), the Optical Monitor was switched off throughout the campaign. The raw data were processed with the Scientific Analysis System (\sas) version 5.4.1. For details on the \xmm\ observations and on the data processing, we refer to the previous work on \hda\ -- the central target of the FOV --  by \citet{SSG04}.

For the purpose of scientific analysis, we adopted a circular extraction region with a radius of 13.2\,arcsec and centered on \cpd. This radius corresponds to half the distance to the nearest neighbouring X-ray source. 
Using the \sas\ task {\sc calview}, we estimated that, at the position of \cpd, the adopted extraction region corresponds to an encircled energy fraction of about 64\% and 63\% respectively for the \mos1 and \mos2 instruments.
Unfortunately, due to the crowded nature of the \ngc\ cluster core in the X-rays (see Fig.~\ref{fig: fov}), the background could not be evaluated in the immediate vicinity of \cpd, but had to be taken from the very few source free regions in the cluster core. We adopted three circular background regions -- labelled A, B and C on Fig.\,\ref{fig: fov} -- centered on $(\alpha, \delta) = (16^\mathrm{h}54^\mathrm{m}31\fs43, -41\degr45\arcmin42\farcs2)$, $(16^\mathrm{h}54^\mathrm{m}23\fs28, -41\degr46\arcmin14\farcs7)$ and $(16^\mathrm{h}53^\mathrm{m}44\fs12, -41\degr53\arcmin34\farcs6)$, and with respective radii of 20, 20 and 25 arcsec. These regions are somewhat offset from the source region but all three are located on the same CCD detector (CCD \#1) as \cpd.

\begin{table*}
\centering
\caption{Time and orbital phase (according to the ephemeris of Table \ref{tab: orbit}) at mid-exposure for each \xmm\ observation of \cpd. The other columns yield the count rates (in units of 10$^{-3}$\,\cnts)  over different energy bands (expressed in keV) for the two \mos\ instruments, as obtained using the \sas\ task {\it emldetect} \citep[see details in][]{SGR05}. The observations lasted on average for 30\,ks (corresponding to a phase interval of 0.14). Note that due to background flares, part of some observations had to be discarded. }
\begin{tabular}{ c c c c c c c c c c c}
\hline
\hline
Obs. & JD & $\psi$ & \multicolumn{4}{c}{MOS1}                                           & \multicolumn{4}{c}{MOS2}\\
\# &$-$2\,450\,000&   & [0.5-10.0]     & [0.5-1.0]     & [1.0-2.5]     & [2.5-10.0]     & [0.5-10.0]     & [0.5-1.0]     & [1.0-2.5]     & [2.5-10.0]   \\
\hline
1 & 2158.214 & 0.819 & $16.5 \pm 0.9$ & $7.7 \pm 0.6$ & $ 8.7 \pm 0.6$ & $0.1 \pm 0.1$ & $14.3 \pm 0.8$ & $ 7.7 \pm 0.6$ & $ 6.2 \pm 0.5$ & $0.4 \pm 0.2$ \\
2 & 2158.931 & 0.113 & $29.7 \pm 1.5$ & $9.8 \pm 0.8$ & $17.0 \pm 0.9$ & $2.9 \pm 0.5$ & $29.0 \pm 1.5$ & $10.1 \pm 0.9$ & $16.6 \pm 1.1$ & $2.2 \pm 0.5$ \\
3 & 2159.796 & 0.468 & $22.8 \pm 1.0$ & $9.0 \pm 0.6$ & $12.2 \pm 0.6$ & $1.6 \pm 0.3$ & $23.6 \pm 1.0$ & $ 9.2 \pm 0.6$ & $13.1 \pm 0.8$ & $1.3 \pm 0.3$ \\
4 & 2160.925 & 0.930 & $19.0 \pm 1.0$ & $8.4 \pm 0.7$ & $ 9.5 \pm 0.7$ & $1.1 \pm 0.3$ & $19.3 \pm 1.1$ & $ 9.3 \pm 0.7$ & $ 9.1 \pm 0.7$ & $1.0 \pm 0.3$ \\
5 & 2161.774 & 0.278 & $19.7 \pm 1.0$ & $8.5 \pm 0.6$ & $10.0 \pm 0.7$ & $1.2 \pm 0.3$ & $21.1 \pm 1.0$ & $ 8.1 \pm 0.6$ & $11.4 \pm 0.7$ & $1.6 \pm 0.3$ \\
6 & 2162.726 & 0.668 & $18.9 \pm 0.9$ & $9.5 \pm 0.6$ & $ 8.9 \pm 0.7$ & $0.5 \pm 0.2$ & $20.3 \pm 1.0$ & $ 9.2 \pm 0.6$ & $10.4 \pm 0.7$ & $0.7 \pm 0.2$ \\
\hline
\end{tabular}
\end{table*}

\begin{figure*}
   \centering
   \includegraphics[width=17cm]{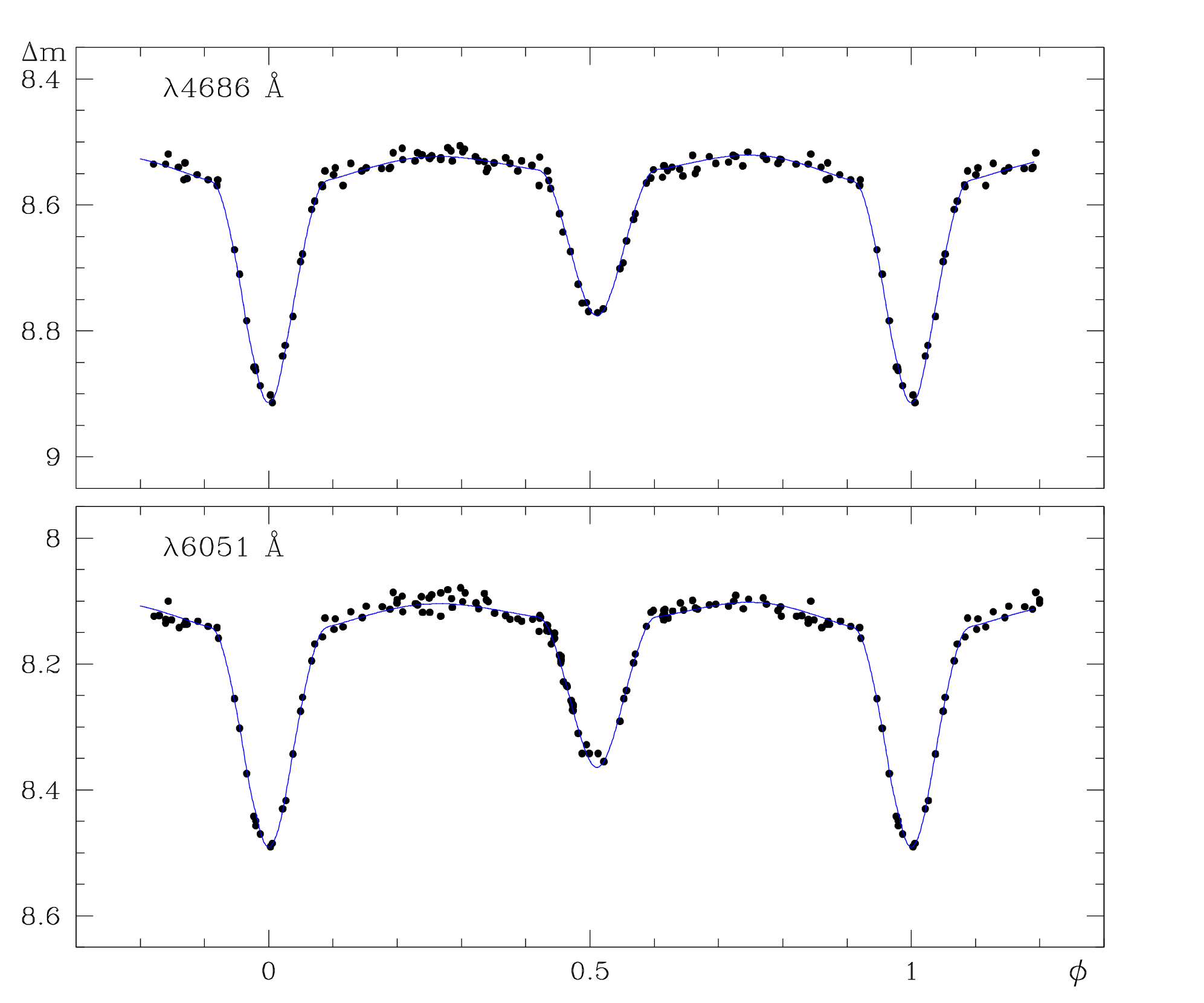}
   \caption{Observed (dots) and modelled (plain lines) light curves at $\lambda 4686$\,\AA ~ and $\lambda 6051$\,\AA. The parameters of the adopted model are presented in Table~\ref{tab: Sol_Par}. The minimum of the primary eclipse is adopted as phase $\phi=0.0$ which, according to the ephemeris of Table \ref{tab: orbit}, corresponds to $\psi\approx0.85$.}
   \label{fig: lcdef}
\end{figure*}

\begin{figure}
   \centering
   \includegraphics[width=8cm]{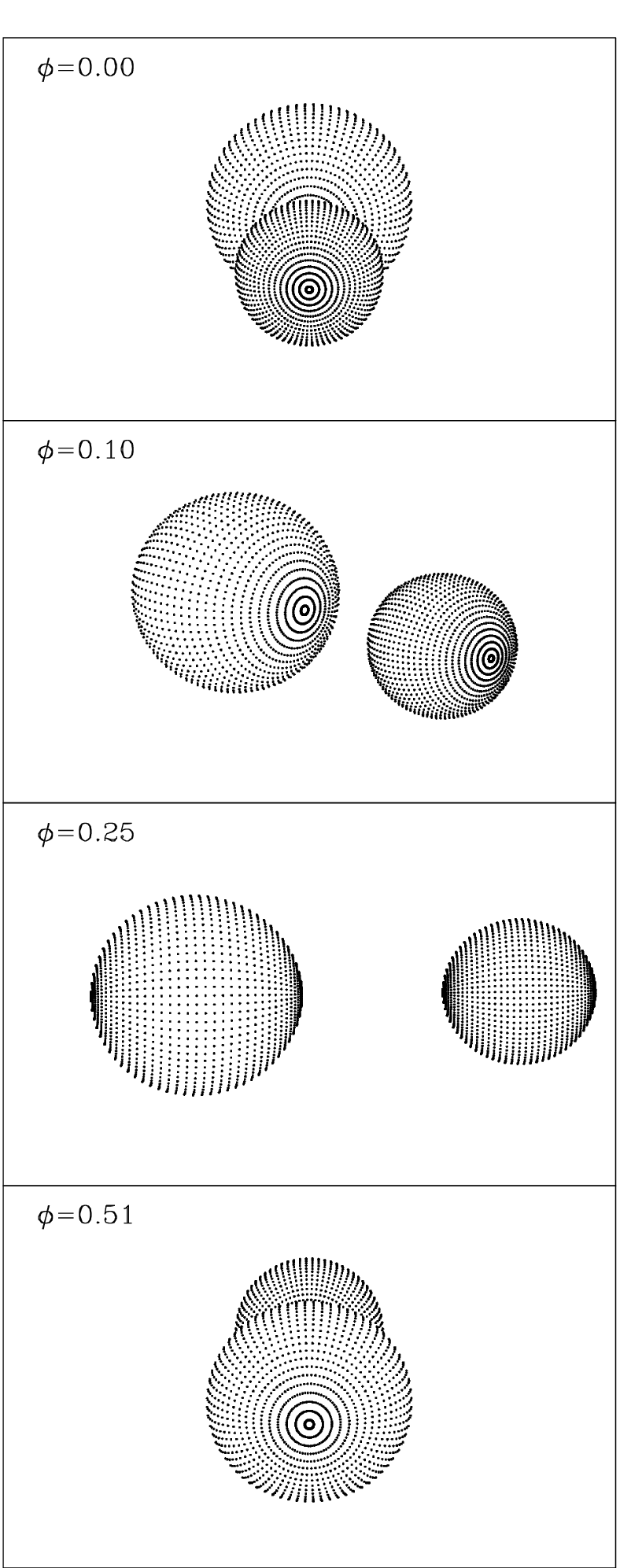}
   \caption{The  model of \cpd\ viewed at different orbital phases $\phi$. The corresponding phases using the ephemeris of Table \ref{tab: orbit} are, from top to bottom, $\psi=0.85$, 0.95, 0.10 and 0.36.
}
   \label{fig: cpdmodel}
\end{figure}

Using the average count rates in each pointing, we built raw and background-corrected broad-band light curves in the range [0.5-10.0\,keV]\footnote{Expressed in pulse-invariant (PI) channel numbers and considering that one PI approximately corresponds to 1\,eV, the adopted range is actually PI$\in$ [500-10\,000].}. We also extracted light curves in three different energy bands: a soft (S$_\mathrm{X}$) band [0.5 - 1.0\,keV], a medium (M$_\mathrm{X}$) band [1.0 - 2.5\,keV] and a hard (H$_\mathrm{X}$) band [2.5 - 10.0\,keV]. 
For comparison, we used the background corrected count rates in each pointing as given in the cluster X-ray source catalogue. These latter values were obtained by means of a psf-model fit to the source using the \sas\ task {\it emldetect} and a spline  background function  \citep[see details in][]{SGR05}. While the catalogue count rates turn out to be about 50\% larger than the extracted count rates, both are in excellent agreement when these latter are corrected for the encircled energy fraction. The obtained X-ray light curves show clear variability. To increase our time resolution, we extracted light curves with temporal bins of 5\,ks, over the same energy ranges as stated above. These latter curves were corrected for the various good time intervals that result from the data processing; they will be discussed in Sect. \ref{sect: X}.

Finally, adopting the same source and background regions, we extracted X-ray spectra for each observation and for each of the two \mos\ instruments. For this purpose, we used the redistribution matrix files ({\it rmf}) provided by the \xmm\ instrument teams and we built the appropriate ancillary response files ({\it arf}) with the help of the \sas\ software. The spectra were binned in such a way as to have at least 10 counts per energy bin. Using the {\sc blanksky} files for the \mos\ instruments, we extracted the spectra corresponding to the adopted source and background regions. The impact of the offset in the background regions on the background spectrum, and on the instrumental emission lines in particular, was found to be negligible.\\

                                                          
 \begin{table}
 \caption{Input parameters for the program of synthesis of the light
         curves.}
\centering
 \label{Input_Par}
 \begin{tabular}{ll}
 \hline
\hline
 \noalign{\smallskip}
 Parameters & Description  \\
 \noalign{\smallskip}
 \hline
 \noalign{\smallskip}
                                                          
 $q=M_1/M_2$      & Mass ratio \\
 $e$              & Eccentricity \\
 $\omega$         & Longitude of periastron of primary star \\
 $F_1, F_2$       & Ratio of surface rotation rate to synchronous \\
                  & rotation rate for both stars \\
 $i$              & Orbital inclination \\                
 $\mu_1$, $\mu_2$ & Roche lobe filling coefficients, $\mu=R/R^*$, where \\
                  & $R$ and $R^*$  are the polar radii for partial and \\
                  & complete filling of  the critical Roche lobe at \\
                  & periastron position ($0 < \mu \le 1$) \\                  
 $T_1$, $T_2$     & Average effective temperatures of the
                    components \\
 $\beta_1$, $\beta_2$ & Gravity darkening coefficients (the temperature of \\ 
                  & an elementary surface area $T=T_{1,2}\times{({g\over{<g>_{1,2}}})}^{\beta_{1,2}}$,\\
                  & where $g$ and $<g>$ are the local and mean gravity \\
                  &  accelerations) \\
 $x_{1,2}$, $y_{1,2}$     & Limb darkening coefficients (see the text) \\
 $A_1$, $A_2$     & Bolometric albedos (coefficients of reprocessing \\
                  & of the emission of a companion by ''reflection'') \\
 $l_3$            & Third light \\
$\Delta \phi$     & Phase shift between the times of conjunction $t_0$ and \\
                  & of periastron passage $T_0$ (Table \ref{tab: orbit})\\
$t_0$             & Time of primary eclipse minimum  \\
 \noalign{\smallskip}
 \hline
 \end{tabular}
 \end{table}



\section{Optical light curve analysis\label{sect: lc}}

Photometric light curves were analysed within the framework of the Roche
model for an eccentric orbit, similar to Wilson's \citep{Wil79} model. The
algorithm is described in detail by \citet{Ant88,Ant96},
here we only briefly describe its main features. The computer code allows
one to calculate a radial velocity curve, the monochromatic light curves and
absorption line profiles of stars simultaneously, either for a circular or
an eccentric orbit. Axial rotation of the stars may be non synchronized with
the orbital revolution. Following \citet{Wil79}, we assumed that the
shapes of the stars coincide with equipotential surfaces in the Roche model
at all orbital phases and both stars retain constant volumes during their
orbital revolution. The tidally distorted surfaces of the stars are heated
by mutual radiation. The intensity of the radiation coming from an
elementary area of the stellar surface and its angular dependence are
determined by the temperature of the star, gravitational darkening, limb
darkening, and heating by radiation from the companion. The input parameters
of the model are summarized in Table~\ref{Input_Par}.

For light curve solution, we fixed some parameters whose values were defined
in previous investigations of the system or can be assumed from global
stellar properties. Namely, we used the known spectroscopic value of mass
ratio $q=M_1/M_2=1.803$, deduced from the data on He I lines 
\citepalias{SHRG03}. A light curve solution is only sensitive to the
temperature ratio between the stars, thus the temperature of one star
should be fixed. Usually it is the more reliably determined temperature of
the primary star. The spectral types of the stars O9\,III (primary) and B1\,III
(secondary) were derived in \citetalias{SHRG03}, but we
pointed out that adopting a main sequence luminosity class for both components
solves much of the inconsistency between the luminosity class III hypothesis and the typical luminosities and radii of giant stars. Our preliminary light curve
solution resulted in stellar radii also suggesting the luminosity class V
for both stars, thus we fixed the average effective temperature of the
primary $T_1=34\,000$~K corresponding to an O9\,V star \citep{HM84}. 
This value is also very close to the one given by the new effective 
temperature scale of O-type dwarfs by \citet{MSH02}.


 \begin{table*}
\centering
 \caption{\cpd\ physical and orbital parameters as obtained from the optical light curve analysis. Two kinds of error estimates are given. The first one defines the confidence intervals inside which the model is still accepted at the 1\% significance level (see text). The second one (given in brackets) corresponds to the 1-$\sigma$ intervals used to define the domain where the true parameter values are expected to lie.}
 \label{tab: Sol_Par}
 \begin{tabular}{lllll}
 \hline
\hline
 \noalign{\smallskip}
 Parameters & $\lambda 4686$~\AA & $\lambda 6051$~\AA & Simultaneous  & Parameter \\
  &  &  & solution & status\\
 \noalign{\smallskip}
 \hline
 \noalign{\smallskip}

 $q=M_1/M_2$ & 1.803                  & 1.803                  & 1.803                  & adopted  \\
 $i$         & $77\fdg35 \pm 0\fdg05$ (0\fdg8) & $77\fdg37 \pm 0\fdg05$ (0\fdg8) & $77\fdg35 \pm 0\fdg05$ (0\fdg8) & adjusted \\ 
 $e$         & $0.020 \pm 0.001$      (0.006)  & $0.020 \pm 0.001$      (0.006)  & $0.020 \pm 0.001$      (0.006)  & adjusted \\
 $\omega$    & $33\degr \pm 8\degr$   (19\degr)& $33\degr \pm 8\degr$   (19\degr)& $33\degr \pm 8\degr$   (19\degr)& adjusted \\
 $\mu_1$     & $0.782 \pm 0.004$      (0.037)  & $0.784 \pm 0.004$      (0.037)  & $0.783 \pm 0.004$      (0.037)  & adjusted \\
 $\mu_2$     & $0.748 \pm 0.003$      (0.050)  & $0.751 \pm 0.003$      (0.050)  & $0.749 \pm 0.003$      (0.050)  & adjusted \\
 $T_1$ (K)   & $34\,000$                       & $34\,000$                       & $34\,000$                       & adopted  \\
 $T_2$ (K)   & $26\,280 \pm 150$      (420)    & $26\,230 \pm 150$      (420)    & $26\,260 \pm 150$      (420)    & adjusted \\
 $L_1/(L_1+L_2)$ & $0.7380$           & $0.7308$               & $0.7379\ |\ 0.7314$    & computed \\
 $L_2/(L_1+L_2)$ & $0.2620$           & $0.2692$               & $0.2621\ |\ 0.2686$    & computed \\
 $F_1$       & 1.0     & 1.0     & 1.0       & adopted\\
 $F_2$       & 1.0     & 1.0     & 1.0       & adopted\\
 $\beta_1$   & 0.25    & 0.25    & 0.25      & adopted\\
 $\beta_2$   & 0.25    & 0.25    & 0.25      & adopted\\
 $A_1$       & 1.0     & 1.0     & 1.0       & adopted\\
 $A_2$       & 1.0     & 1.0     & 1.0       & adopted\\
 $l_3$       & 0.0     & 0.0     & 0.0       & adopted\\
 $x_1$       & $-0.213$& $-0.188$                          &           $-0.213\ |\ -0.188$  & adopted \\
 $y_1$       & \hspace*{2mm}0.724   & \hspace*{2.2mm}0.643   &\hspace*{1.3mm} $0.724\ |\ \hspace*{3.3mm}0.643 $    & adopted \\
 $x_2$       & $-0.124$& $-0.112$                          &           $-0.124\ |\ -0.112$  & adopted \\
 $y_2$       & \hspace*{2mm}0.663   & \hspace*{2.2mm}0.559   &\hspace*{1.3mm} $0.663\ |\ \hspace*{3.3mm}0.559 $    & adopted \\
 $\Delta\,\phi$ & $0.1537 \pm 0.0007$ (0.0011) & $0.1537 \pm 0.0007$  (0.0011) & $0.1537 \pm 0.0007$  (0.0011) & adjusted\\
 $t_0\ (HJD-2\,450\,000$)& 2399.909 & 2399.909 & 2399.909 & computed \\
 \noalign{\smallskip}
 \hline
 \noalign{\smallskip}
 Relative radii ($R/a$) & & \\
 \noalign{\smallskip}
 \hline
 $r_1(pole)$  & $0.3127 \pm 0.0016$ (0.0148) & $0.3135 \pm 0.0016$ (0.0148) & $0.3131 \pm 0.0016$ (0.0148) \\
 $r_1(point)$ & $0.3351 \pm 0.0022$ (0.0203) & $0.3362 \pm 0.0022$ (0.0203) & $0.3357 \pm 0.0022$ (0.0203) \\
 $r_1(side)$  & $0.3205 \pm 0.0018$ (0.0164) & $0.3214 \pm 0.0018$ (0.0164) & $0.3210 \pm 0.0018$ (0.0164) \\
 $r_1(back)$  & $0.3290 \pm 0.0020$ (0.0182) & $0.3300 \pm 0.0020$ (0.0182) & $0.3295 \pm 0.0020$ (0.0182) \\
 $r_2(pole)$  & $0.2268 \pm 0.0009$ (0.0152) & $0.2277 \pm 0.0009$ (0.0152) & $0.2271 \pm 0.0009$ (0.0152) \\
 $r_2(point)$ & $0.2421 \pm 0.0012$ (0.0210) & $0.2433 \pm 0.0012$ (0.0210) & $0.2425 \pm 0.0012$ (0.0210) \\
 $r_2(side)$  & $0.2306 \pm 0.0010$ (0.0163) & $0.2316 \pm 0.0010$ (0.0163) & $0.2309 \pm 0.0010$ (0.0163) \\
 $r_2(back)$  & $0.2384 \pm 0.0011$ (0.0189) & $0.2395 \pm 0.0011$ (0.0189) & $0.2387 \pm 0.0011$ (0.0189) \\

 \noalign{\smallskip}
 \hline
 \end{tabular}
 \end{table*}


Gravity-darkening coefficients $\beta_1=\beta_2=0.25$ and albedos
$A_1=A_2=1$ were assumed as typical for early type stars. We used the nonlinear
`square-root' limb-darkening law \citep{DCG92, DCCG95, vHa93}:
$I(\cos\gamma)=I(1)[1-x(1-\cos\gamma)-y(1-\sqrt{\cos\gamma})]$, where
$\gamma$ is the angle between the line of sight and the normal to the
surface, $I(1)$ is the intensity for $\gamma = 0$, and $x,y$ are the limb
darkening coefficients. As shown by \citet{vHa93}, this is the
most appropriate limb-darkening law at optical wavelengths for
$T\geq10\,000$~K. The rotation of both stars is assumed to be synchronous
with the orbital one $F_1=F_2=1$.

The adjustable parameters of the model were the following:
the Roche lobe filling coefficients for the primary and secondary
$\mu_1,\mu_2$ (calculated for the time of periastron passage),
the average effective temperature of the secondary star $T_2$, the orbital
inclination $i$, the eccentricity $e$, the longitude of periastron of the primary
$\omega$. While doing minimization, every model light curve was also shifted
along the magnitude axis until the best fit between the model and observed
curves was achieved.

Initial phases $\psi$ of observational data points were calculated using the
spectroscopic ephemeris of Table \ref{tab: orbit}:
$HJD = 2\,452\,400.284 + 2{\fd}44070 \times E.$ Since our model assumes an
orbital phase $\phi$ equal to zero at the time of conjunction (the secondary star
being in front), the observed light curve was then shifted in phase
by $\Delta\,\phi$, according to $\psi=\phi-\Delta\phi$. 
The value of $\Delta\,\phi$ was determined by the minimum of
the deviation between the observed and model light curves.


The estimation of adjustable parameters was done with the well-known simplex
algorithm (Nelder and Mead's method) \citep{Him71, KL87}. In the vicinity of the minima found, additional
calculations were done on a fine grid, in order to explore the details in shape of
the deviation surface and to estimate the errors on the
parameters.  The resulting parameters for the solutions corresponding to \l4686\AA, to \l6051\AA\ and to the simultaneous adjustment at both wavelengths are presented in Table \ref{tab: Sol_Par}. Two kinds of confidence intervals have been computed and are also  given in Table \ref{tab: Sol_Par}.
The first one corresponds to a test of the adequacy of the model.
The confidence intervals for the parameters are estimated using an absolute critical value of $\chi^2$ corresponding to a significance level of 1\%. This first approach rather defines the zones of variation of the parameters that still lead to an acceptation of the model. The obtained error bars  are rather small.
The second kind of confidence intervals corresponds to a critical value which is defined relatively to the obtained minimum $\chi^2$ of the fit, increased by a value corresponding to a significance level of 0.1\%. This latter interval corresponds to a 3-$\sigma$ deviation and it has been transformed to a 1$\sigma$-uncertainty in the sake of coherence with the radial velocity adjustment. This approach is reminiscent to a search for the zone where lie the true values of the parameters.

 Figure \ref{fig: lcdef} exhibits the observed light curves corresponding to
$\lambda 4686$\,\AA\ and to $\lambda 6051$\,\AA\ along with the model predictions of the simultaneous solution. The final model
for \cpd\ viewed at different orbital phases is presented in Fig.~\ref{fig: cpdmodel}.

\section{\cpd\ orbital and physical parameters \label{sect: discuss_photo}}

\subsection{Period $P$ \label{ssect: discuss_period}}
Since the time base of our photometric campaign is {\it only} 28 days long, it provides little constraint on the period. Indeed the width of the associated peak in the periodogram is about $3.6\times 10^{-2}$\,d$^{-1}$, yielding an uncertainty of about  $2.1\times 10^{-2}$\,d (corresponding to one tenth of the peak width) on the value of a period determined from the photometric set only. As a consequence, we choose to keep the period fixed at the value determined from the much longer time span of our spectroscopic data set. We thus retain  $P=2.44070$\,d for \cpd.


 \begin{table}
\centering
 \caption{\cpd\ absolute parameters. The errors on the luminosities and the magnitudes were estimated assuming a formal error of 1000\,K on the temperatures.}
 \label{tab: Abs_Par}
 \begin{tabular}{lcc}
 \hline
\hline
 \noalign{\smallskip}
 Parameters & Primary & Secondary  \\
 \noalign{\smallskip}
 \hline
 \noalign{\smallskip}

 $a(R_{\sun})$                   & \multicolumn{2}{c}{$23.18\pm0.18$}\\
 $R(R_{\sun})$                   &  $7.45\pm0.45$  &  $5.39\pm0.43$  \\
 $M(M_{\sun})$                   & $17.97\pm0.45$  &  $9.96\pm0.22$  \\
 $T(K)$                          & 34\,000         &  26\,260        \\
 $\log(L_\mathrm{bol}/L_{\sun})$ & $4.82\pm0.07$   &  $4.09\pm0.10$  \\
 $\log(g)$                       & $3.93\pm0.48$   &  $3.96\pm0.64$  \\
 $M_\mathrm{V}$                  & $-4.00\pm0.21$  &  $-2.98\pm0.31$ \\

 \noalign{\smallskip}
 \hline
 \end{tabular}
 \end{table}

\subsection{Eccentricity $e$ \label{ssect: discuss_ecc}}
The values of the eccentricity obtained from the analysis of the light curve and of the radial velocity curve are in excellent agreement. From our data, the separation between the two light minima is indeed clearly different from half an orbital cycle and the \cpd\ orbit is thus slightly  eccentric. 

Recently, \citet{StB04} led a photometric campaign searching for new variables in \ngc. Using the period from \citetalias{SHRG03}, they obtained independent light curves for \cpd\ in the Str\"omgren system.  Surprisingly, their data set reveals almost perfectly symmetric light curves with the two light minima separated by exactly half a cycle, thus indicating either a non eccentric system or a longitude of periastron very close to  90 or 270\degr. No detailed analysis of the light curve has been published yet, but the differences between the \citeauthor{StB04} observations and ours are quite intriguing. 

In our data, the ingress of the secondary eclipse has been observed during three different nights spread over the one month run. It is therefore well defined and clearly indicates a slight eccentricity, except if some systematic biases were present. Our observing run lasted for 28 days and the \cpd\ light curves displayed in Fig.~\ref{fig: lcdef} show smooth ellipsoidal variations and well behaved eclipses. Spread over at least two years and acquired more recently, the \citeauthor{StB04} data set is larger, especially in the $y$ and $b$ bands, though with some gaps in the phase coverage. Their published light curves display several striking features. First, the primary eclipse seems to  vary over the time: it presents different depths over different cycles and shows different ingress and egress shapes. Rapid variations  are also observed slightly before the primary eclipse as well as slightly after the secondary one. The right wing of the secondary eclipse displays an inflection point in the $y$ and $b$ bands, while  a strange {\it bifurcation} is observed in the $u$ band. Finally, even outside the eclipses, the behaviour of the system is clearly not as quiet as in our data set (see Fig.~\ref{fig: lcdef}). 

While a change of the period or of the eccentricity with time is hard to explain, a change in the longitude of the periastron  could mimic a non-eccentric system. Another hypothesis, also mentioned by \citeauthor{StB04}, is that the observed dispersion of their light curves reveal the signature of some kind of activity in \cpd. Under this hypothesis, the system could have remained in a quiet state during the 11-cycle duration of our observations, while \citeauthor{StB04}\ could have observed different activity states during the longer time-span of their campaign. 

\subsection{Longitude of periastron $\omega$ \label{ssect: discuss_omega}}
The two values for the primary longitude of periastron obtained from the spectroscopy ($\omega=149\degr$) and  from the photometry ($\omega=33\degr$) are clearly not consistent. In \citetalias{SHRG03}, we also computed an orbital solution including all published primary RVs and we obtained an argument $\omega=27\pm31\degr$ closer to the latter photometric value. In principle, the light curve analysis is a more powerful tool to derive accurate values for $\omega$, again from the separation between the two light minima. In Fig.~\ref{fig: lcdef}, the separation between the primary and secondary eclipses is slightly larger than half a cycle. This indicates that the longitude of periastron  is located between 0\degr\ and 90\degr, thus rejecting the much larger spectroscopic value.

In Paper I, we already noted the large dispersion in the values deduced from data sets associated with different lines, ranging from $\omega=99\degr$ to 190\degr. We tentatively suggested that this was linked to the difficulty to accurately determine the  periastron  argument in such a slightly eccentric system. From our orbital solution, we however derived a reasonable error bar of 10\degr. While searching for the origin of the discrepancy  between the photometric and spectroscopic solutions, we have investigated this point more deeply. Adopting the orbital parameters of Table \ref{tab: orbit}, we computed a set of orbital solutions, varying the periastron argument from 0\degr\ to 360\degr. The obtained curves are very similar in shape; the main difference is a shift in radial velocity of an amplitude of about 10\,\kms\ peak-to-peak. Comparing this with the root-mean-square (r.m.s.) residual of 4.8\,\kms\ of our orbital solution  gives us a first impression that the periastron argument is probably loosely constrained by the radial velocity solution and that the quoted error-bar could be underestimated in this particular case.

In a second approach, we performed Monte-Carlo simulations adopting a Gaussian distribution of the errors on the measured radial velocities (RVs). For the primary, we adopted a standard deviation of 4.8\,\kms, thus equal to the r.m.s. residual of our fit. For the secondary component, we accounted for the obtained ratio between the primary and secondary uncertainties, $s_y/s_x=2.1$, as quoted in \citetalias{SHRG03}. Finally, for each observation, we scaled the dispersion according to the relative weighting adopted to compute the orbital solution. For each measured RVs, we randomly drew a series of 10\,000 simulated RV points from these distributions, so building an equivalent number of simulated data sets. We then computed the corresponding orbital solutions using the same method as the one described in \citetalias{SHRG03}. We finally computed the distributions of the resulting orbital elements. This latter approach allows to estimate the errors assuming a random dispersion of the observed points. This evidently does not account for possible systematic errors or outstanding points.

\begin{figure}
\centering
\includegraphics[width=\columnwidth]{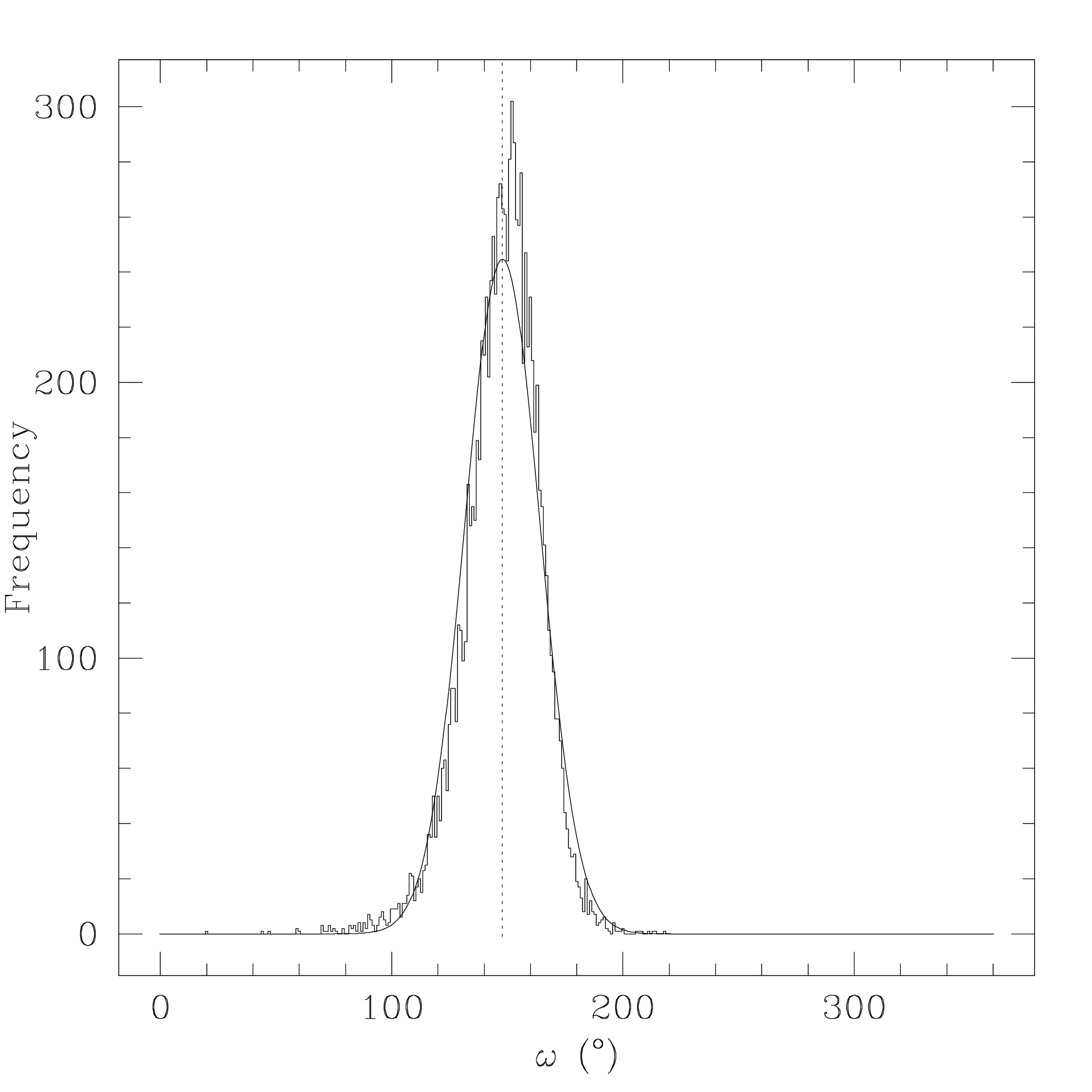}
\caption{Distribution of the longitude of periastron $\omega$ in a set of 10\,000  simulated orbital solutions built using Monte-Carlo techniques (see text). An equivalent Gaussian characterized by the same estimated mean and standard dispersion and with an equivalent surface has been overplotted. }
\label{fig: omega}
\end{figure}

We found that all the orbital parameters follow a Gaussian distribution, centered on the values of Table \ref{tab: orbit}, except the longitude of periastron (and thus the time of periastron passage). The simulated 1-$\sigma$ dispersions were found to be systematically, but not dramatically, higher than the published uncertainties. The difference is, on average, not larger than 80\% but can reach a factor of 3. These new values for the uncertainties are quoted in the right column of Table \ref{tab: orbit}. Concerning the distribution of the periastron argument, Fig.~\ref{fig: omega} shows that it significantly deviates from a Gaussian distribution. The width of the peak however does approximately correspond to the width of an equivalent Gaussian characterized by an estimated  standard deviation equal to the one of the simulated distribution and by an equivalent surface. We thus retain the 1-$\sigma$ dispersion of the distribution as a good estimator of the typical error on the determined spectroscopic value for $\omega$. As a consequence, while the quoted error on the periastron argument was indeed underestimated in \citetalias{SHRG03}, this new estimate explains rather well the dispersion observed from time to time but still rules out the photometric value $\omega=33\degr$. The origin of the inconsistency between the photometric and spectroscopic values, as well as with the \citet{StB04} data, should be looked for elsewhere.

One could indeed think of a possible physical effect that would modify the observed RV curve compared to the {\it true} curve of the system. In particular, a modification of the position of the spectral line centroids could produce a different RV value compared to the {\it true} velocity of the stars. In such a slightly eccentric binary as \cpd, it is also plausible that a small variation in the measured RVs could mimic orbits with quite a different periastron argument.  Though the exact nature of the phenomenon is unknown, we tentatively linked it to a possible manifestation of the Barr effect. 

The Canadian amateur astronomer, J.\ Miller Barr noted that the
longitudes of periastron of spectroscopic binaries are not uniformly
distributed between $0^{\circ}$ and $360^{\circ}$ \citep{Barr08}. Out of 
30 spectroscopic binaries with elliptical orbits, apparently only four
had $\omega$ between $180^{\circ}$ and $360^{\circ}$, all others had
their longitude of periastron in the first two quadrants. Barr advanced
two possible explanations for this systematic effect: either the
pressure or temperature effects in the atmospheres of the stars shift
their spectral lines with respect to their genuine orbital motion, or a
non-uniform brightness of the components combined with a large
rotational velocity causes the spectral lines to become asymmetric.
Although Barr included several Cepheid variables in his sample, a
similar effect was (re-)discovered by  \citet[ see also the
discussion by \citealt{Bat83, Bat88}]{Str48}. Struve apparently found an excess of
systems with $\omega$ in the first quadrant. He suggested that this
could be due to streams of gas between the stars which lead to spurious
eccentricities and values of $\omega$ in the first quadrant. The
existence of the Barr effect was confirmed by the studies of \citet{Fra79}
 and \citet{How93}. \citeauthor{Fra79} used the data from the VIIth
Catalogue of orbital elements of spectroscopic binaries and
found a distribution of $\omega$ for systems with large eccentricities
($e \geq 0.6$) that shows an excess of systems with $\omega = 0^{\circ}$
and a flat minimum around $\omega = 250^{\circ}$. The effect was most
prominent in systems with short orbital periods. \citet{How93} analysed
the effect by means of non-parametric statistical tests, restricting his
sample to systems with orbital solutions of reasonable quality. He found
a statistically significant effect only for systems with orbital periods
shorter than 3\,days. The distribution of $\omega$ peaks  at a preferred
direction of $\omega \simeq 100^{\circ}$, corresponding to a shallower,
longer rising branch in the radial velocity curve and a steeper, shorter
falling branch. Howarth interpreted this effect as the result of a gas
stream from the primary towards the secondary, though no simulation of 
the phenomenon has been performed to check its exact influence on the RV curve.

\subsection{Time of periastron passage $T_0$ \label{ssect: discuss_t0}}
The difference in the spectroscopically and photometrically determined times of periastron passage directly results from the inconsistency between the values of the periastron argument derived using the two techniques. This problem has already been extensively described in the previous paragraph (Sect. \ref{ssect: discuss_omega}). We just note here that adopting a periastron argument $\omega =33\degr$ yields a value for the time of periastron passage of $T_0=2\,452\,399.498$ (HJD). \\

\begin{figure}
   \centering
   \includegraphics[width=\columnwidth]{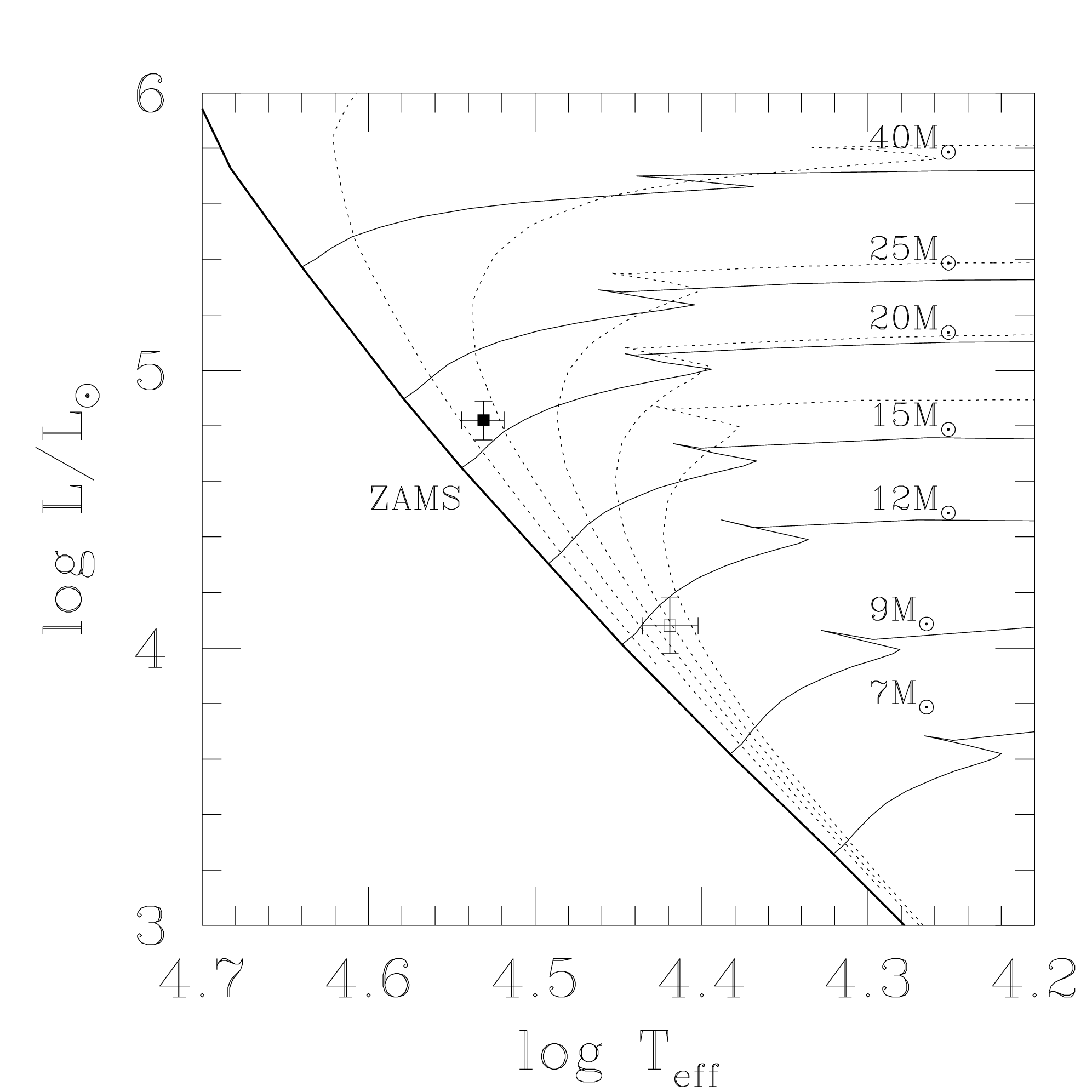}
   \caption{Position of the primary (filled symbol) and secondary (open symbol) components of \cpd\ in the H-R diagram. A formal error of 1000\,K has been adopted on the temperatures. Solid lines: evolutionary tracks from \citet{SSM92} for different initial masses. Dotted lines: isochrones ranging, from left to right, from 2 to 10\,Myr with a step of 2\,Myr.
}
   \label{fig: HR}
\end{figure}

\subsection{\cpd\ physical parameters \label{ssect: discuss_phys_par}}

Thanks to the light curve analysis, the inclination of the system is now very well constrained. Combining this with the spectroscopic information of Table \ref{tab: orbit}, we derived absolute values for the system separation and the star radii and masses. We also derived their luminosity and the surface gravity. The physical parameters of both stars are given in Table \ref{tab: Abs_Par}. With an absolute radius of $R_1 =7.45 \pm 0.45$\,\rsol, the primary component is  slightly smaller than typical O9~V stars. \citet{HP89}, \citet{SK82} and \citet{VGS96} respectively listed radii of 8, 9.5 and 8.8 \rsol. The observed radius is however larger than the typical O9.5~V radius of 7 \rsol\ given by \citeauthor{HP89}. 
Adopting the bolometric correction of \citeauthor{HM84}, $BC=-3.3\pm0.1$, we derived a visual absolute magnitude $M_\mathrm{V,1}=-4.00\pm0.21$, fainter than the values of $-$4.5, $-$4.2, $-$4.5 and $-$4.43 respectively reported by \citet{HM84}, \citet{HP89}, \citet{SK82} and \citet{VGS96}, though again in agreement with the slightly later spectral-type O9.5~V. 
Comparing the obtained values with those of other eclipsing early-type binaries listed by \citet{Gie03} clearly indicates that the physical parameters of the primary in \cpd\ correspond to the observed range for O9 dwarfs. \citet{VCV97} reported a mass of 19~\msol\ for the O9~V component in \object{HD~165921} though with a relatively smaller radius ($R=6.13$~\rsol). On the other hand, the \cpd\ primary is slightly larger and heavier than the O9.5 dwarf components in \object{CPD$-$59\degr2603} \citep[ $M=14.5$~\msol, $R=4.9$~\rsol]{RSA01}, \object{HD\,193611} \citep[ $M=16.6~+~16.3$~\msol, $R=7.4~+~7.4$~\rsol]{PoH91} or \object{HD~198846} \citep[ $M=17.0-17.7$~\msol, $R=5.7-7.7$~\rsol]{SSF94, HiH95, BMM97}. The dwarf nature of the primary star is consistent with the derived surface gravity (altough the corresponding error is rather large).

From the effective temperature calibration of \citeauthor{HM84}, the secondary temperature corresponds to a spectral sub-type B0.5, in rough agreement with the B1 spectral type obtained from spectroscopy. Its radius and visual magnitude however fall within the expected range for B1-2 stars \citep{HM84, SK82}. The secondary is also slightly smaller and lighter than the B1~V component in \object{HD~175514} \citep[ $M=13.5$~\msol, $R=5.9$~\rsol]{BHA87}. All in all, and accounting for the uncertainties on the spectroscopic data, adopting a B1.5~V spectral sub-type for the secondary in \cpd\ yields a better match between its physical parameters and the typical observed and theoretical values expected for such a star.

The locations of the \cpd\ components in the H-R diagram are shown in Fig.~\ref{fig: HR} together with the evolutionary tracks of \citet{SSM92}. A rough interpolation from these tracks yields initial  masses $M_1^{(0)}=23.7$\,\msol\ and $M_2^{(0)}=11.1$\,\msol\ and current ages between 3 and 8~Myr. These ages do well reproduce the range of derived values for the \ngc\ cluster \citep[see the cluster literature review in][]{SGR05}. In such a small time span, the  actual masses of the stars remain close to their initial masses and are thus quite larger than the observed masses of about 18 and 10\,\msol\ (Table \ref{tab: Abs_Par}). 
In a binary system, mass exchange between its components, through e.g. Roche lobe overflow, could alter their evolutionary status compared to single star models. From the photometric light curve, \cpd\ is actually a well detached system. Due to its young age, it is thus very unlikely that the system could have undergone such a phenomenon (now interrupted) in its past history. New evolutionary tracks that account  for the effect of  rotation could help to investigate this apparent discrepancy. Finally, comparing the absolute magnitudes obtained in Table \ref{tab: Abs_Par} with the visual magnitude $V=8.228$ of \cpd\ \citep{SBL98}, we estimated the distance of the object. We adopted a colour excess $E(B-V)=0.49$ and $R=3.3$ as derived by \citet{SBL98}. We finally obtained $DM=10.92\pm0.16$, in excellent agreement with the cluster average distance modulus $DM=11.07\pm0.04$ \citep{SGR05}.

\begin{figure}
\centering
\includegraphics[width=8cm]{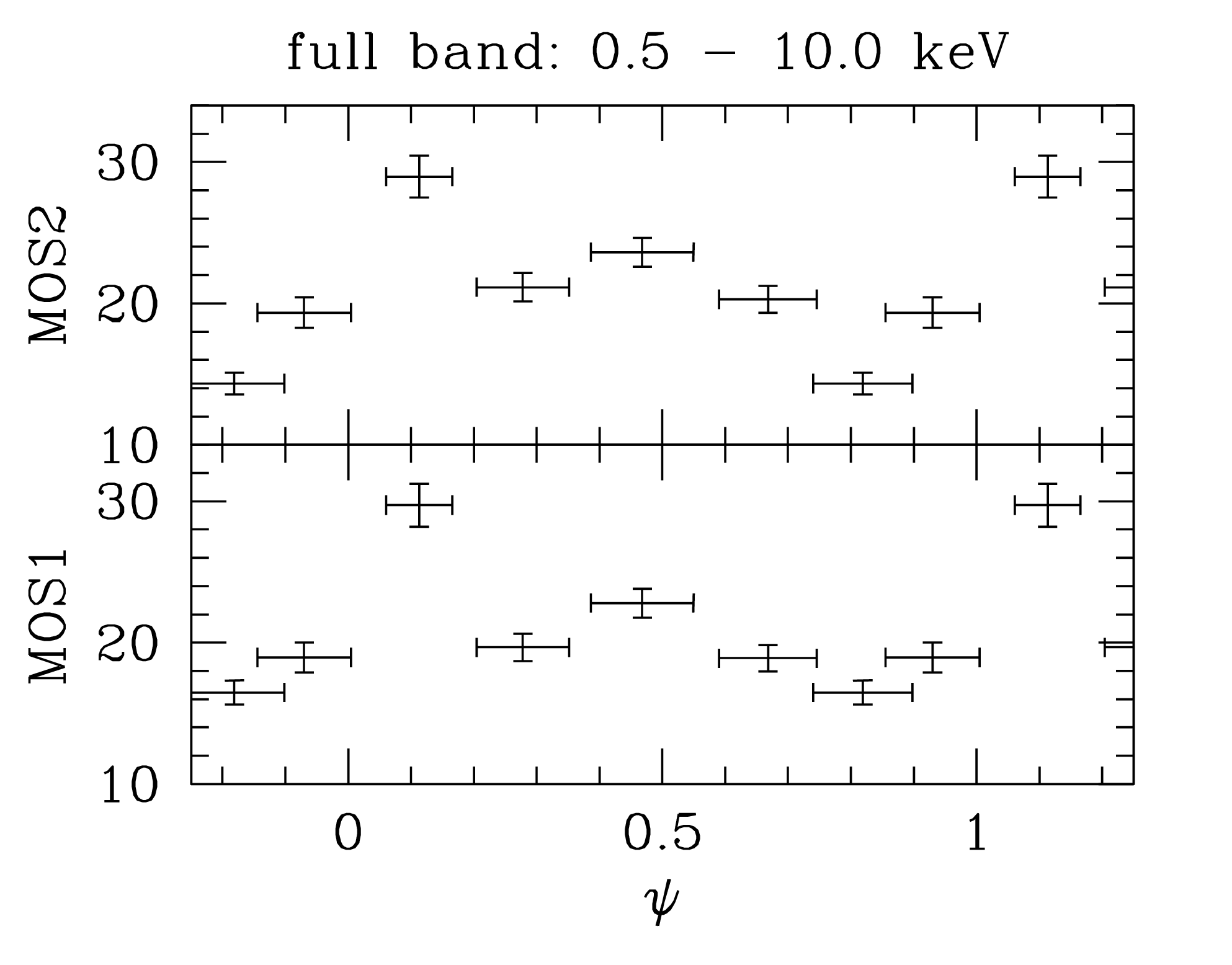}
\caption{Net \epicmos\ count rates of \cpd\ as a function of orbital phase and averaged over the duration of each pointing \citep[from ][]{SGR05}. The vertical axes are in units $10^{-3}$\,\cnts. The horizontal {\it error bars} indicate the extension in phase of the corresponding pointing.}
\label{fig: cpd42_lc}
\end{figure}

\begin{figure}
\centering
\includegraphics[width=8.5cm]{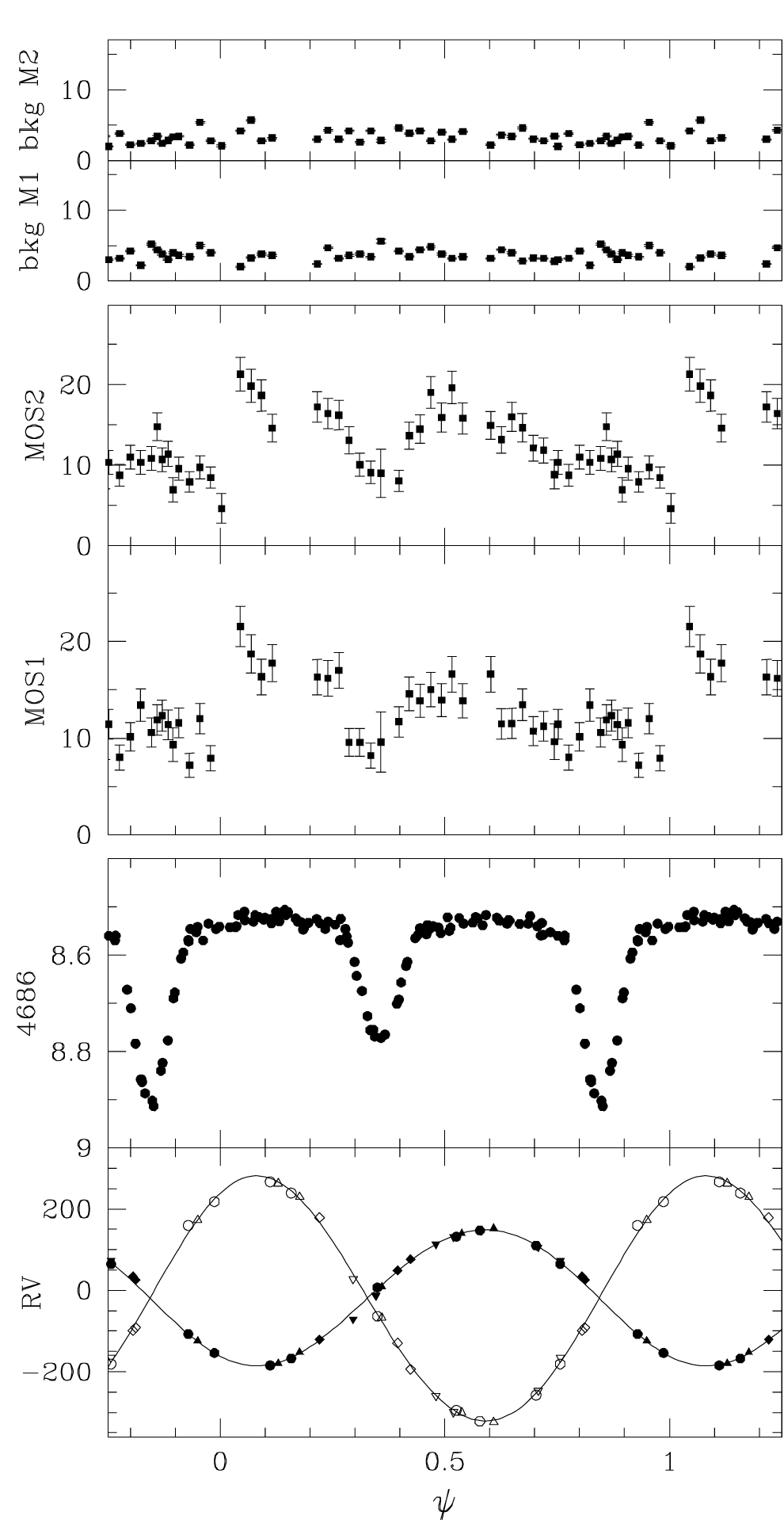}
\caption{{\bf Top panel:} \epicmos\ background count rates in the [0.5-10.0\,keV] band. {\bf Middle panel:} \cpd\  \epicmos\ background-corrected count rates in the same energy range. The time binning of these two panels is 5\,ks. The vertical axes are in units $10^{-3}$\,cnt\,s$^{-1}$. No correction for the limited encircled energy fraction has been applied. {\bf Lower panel:} RV curve (in \kms) and optical light curve (in mag) of \cpd. Note the coincidence of the X-ray drop around $\psi=0.35$ and the time of conjunction with the primary star being in front, as well as the lack of coincidence of the secondary eclipse with the passage at the systemic velocity. }\label{fig: cpd42_lc5ks}
\end{figure}

\section{X-ray light curves and spectral analysis \label{sect: X}}

\begin{figure*}
\centering
\includegraphics[width=5.9cm]{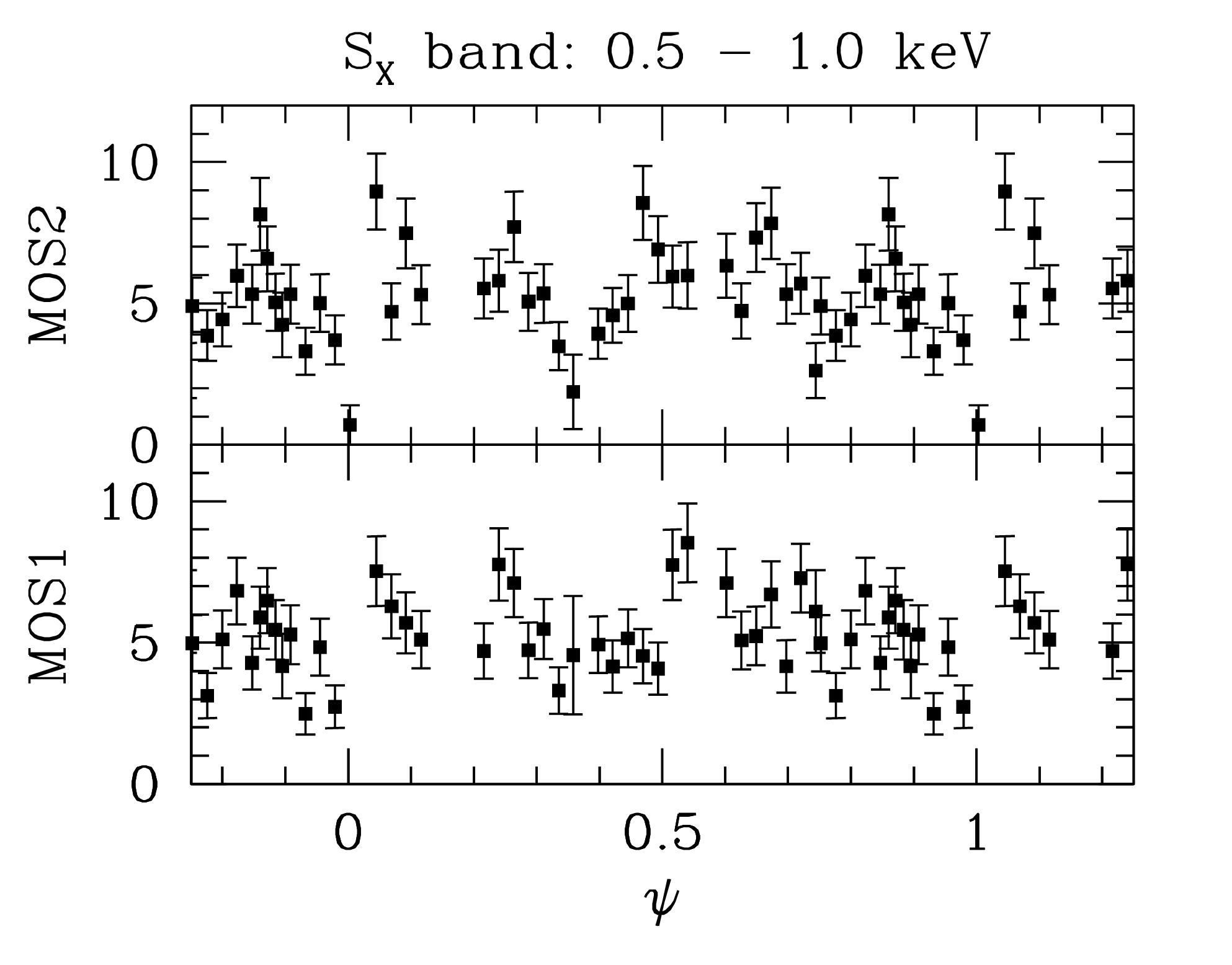}
\includegraphics[width=5.9cm]{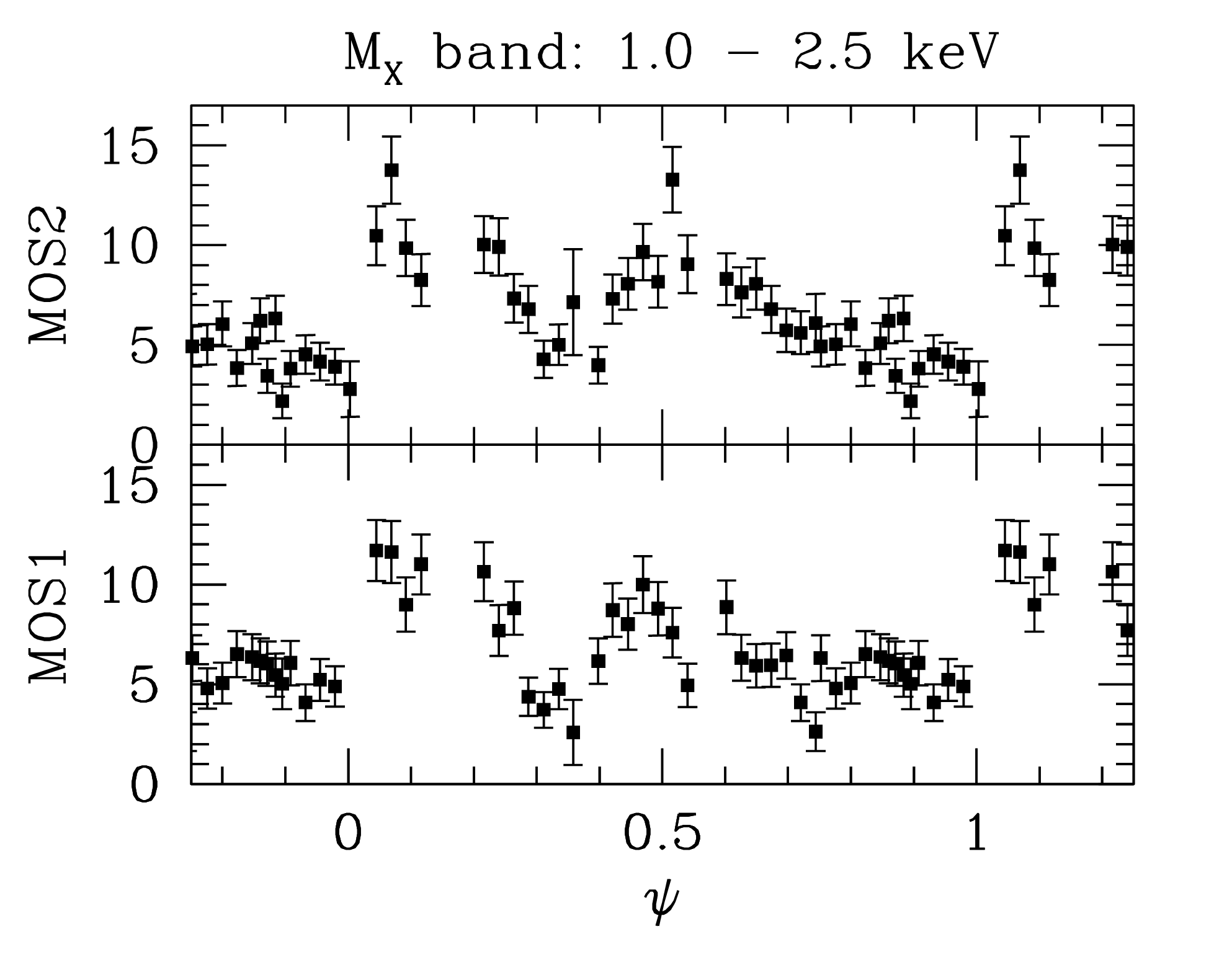}
\includegraphics[width=5.9cm]{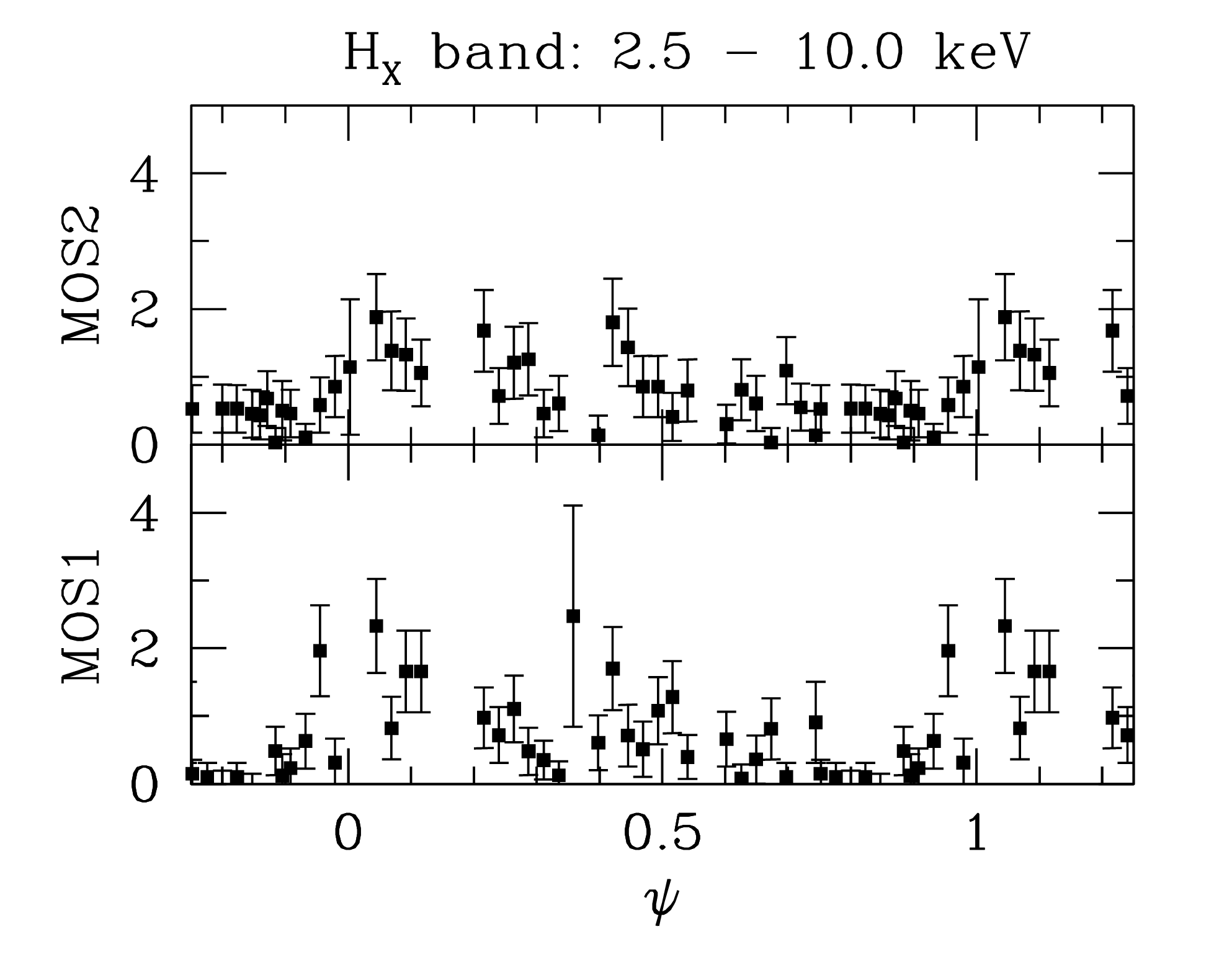}
\caption{\cpd\ \epicmos\ background-corrected count rates in the three energy bands as a function of orbital phase. The time binning is 5\,ks. The vertical axes are in units $10^{-3}$\,cnt\,s$^{-1}$. No correction for the limited encircled energy fraction has been applied.}
\label{fig: cpd42_lc5ks_eb}
\end{figure*}

The X-ray light curves of \cpd\ as seen by the two \mos\ cameras are shown in Fig.\,\ref{fig: cpd42_lc}. The count rates, averaged over the duration of each pointing,  were taken from the \ngc\ X-ray source catalogue of \citet{SGR05} and were obtained using the \sas\ task {\it emldetect}. The count rates are thus corrected for the effects of exposure, vignetting and finite size of the extraction region. They also account for the background subtraction.
It is clear from  Fig.\,\ref{fig: cpd42_lc} that the X-ray emission from \cpd\ displays strong signs of variability. A $\chi^2$ test of hypothesis consistently rejects, at the 1\% significance level, the null hypothesis of constant rates in the [0.5 - 10.0\, keV] band and in the M$_\mathrm{X}$ and H$_\mathrm{X}$ bands. Fig.~\ref{fig: cpd42_lc} 
also indicates that the phase coverage of the orbital cycle is almost complete with only a small gap slightly before phase $\psi=0.2$. To increase our time resolution, we also extracted background-corrected light curves with a time binning of 5\,ks. 
Figure\,\ref{fig: cpd42_lc5ks} shows that the count rate changes by about a factor two over relatively short time scales. These variations are also seen in the different energy ranges (Fig.~\ref{fig: cpd42_lc5ks_eb}) and  are most prominent in the intermediate (M$_\mathrm{X}$) band. As in Fig.~\ref{fig: cpd42_lc5ks}, they suggest a double-peaked light curve with two broad maxima around phases $\psi\approx 0.1$ and 0.5. From the top panels of Fig.~\ref{fig: cpd42_lc5ks}, we conclude that the observed modulations are clearly not due to background fluctuations.  Note that, in Figs.~\ref{fig: cpd42_lc5ks} and \ref{fig: cpd42_lc5ks_eb}, no correction for the limited encircled energy fraction has been applied, neither for the vignetting nor for the exposure. This explains the lower count rates obtained compared to Fig.~\ref{fig: cpd42_lc}.

\begin{figure}
\centering
\resizebox{8cm}{5.cm}{\includegraphics[width=8cm]{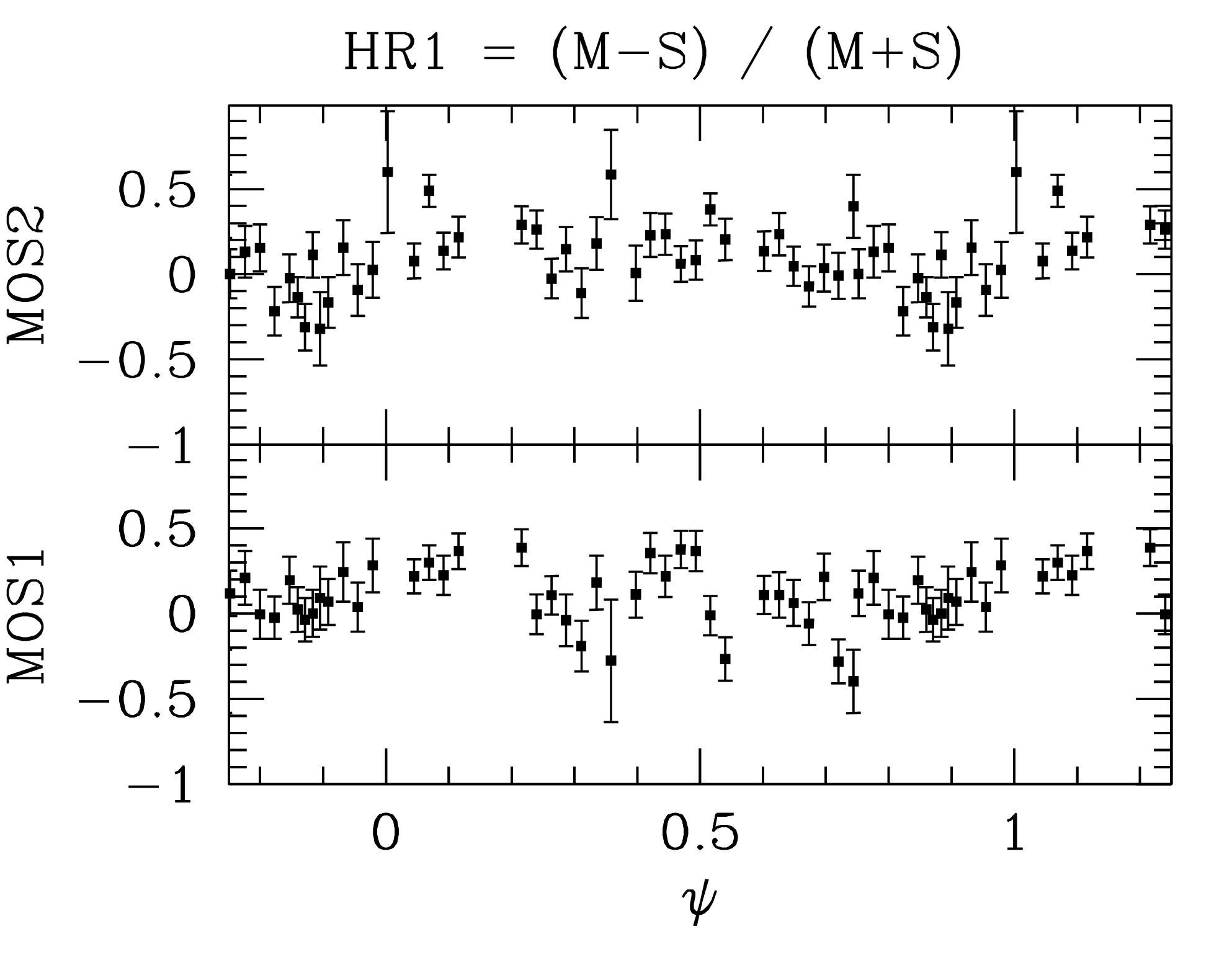}}
\caption{ Hardness ratios vs. phase for the \epicmos\ instruments. The adopted definition for $HR_1$ is given on top of the panel.}
\label{fig: hr}
\end{figure}

One of the main pictures in the X-ray light curve is a sensible decrease of the signal between phase $\psi=0.27$ and $0.45$, which almost exactly corresponds to the time of the secondary minimum in the optical light curve. The observed modulations are probably phase-locked since, for example, the two wings of the {\it eclipse} have been observed during two different orbital revolutions. However, except near $\psi=0.85$, the different pointings do not overlap in phase. One can therefore not definitively assert the phase-locked behaviour of the observed X-ray light curves. 

Figure \ref{fig: cpd42_lc5ks} seems thus to indicate two different emission levels: a higher state between $\psi\approx 0.0$ and 0.5, during which the {\it eclipse} is observed, and a lower state between $\psi\approx0.6$ and 1.0, where no counterpart of the primary eclipse can be seen.
The hardness ratio curves are shown in Fig.~\ref{fig: hr}. Though the error bars are quite large, they  seem to indicate that the emission is slightly softer around phase $\psi=0.3$, so approximately at the time of conjunction, while it is presumably harder at the maximum of the emission. 


While the orbit is presumably not circular, the value of the eccentricity is pretty small and it is unlikely that the variation of the distance between the two stars plays a significant role in \cpd. In consequence, the observed modulations of the  X-ray emission are more probably due to a modification of the line of sight towards the system while it is revolving around its center of mass. The observed X-ray light curves will be discussed in the framework of a wind interaction model presented in the next section (Sect. \ref{ssect: Xinteract}). 

As a next step in the analysis, we attempt to constrain the physical properties of the X-ray emission by adjusting a series of models to the obtained spectra for each pointing. We simultaneously fitted the two \mos\ spectra using the \xspec\ software v.11.2.0 \citep{arn96}.
Using the $B - V$ colours quoted by \citet{BVF99} and \citet{SBL98}, we infer a colour excess of about $E(B - V) = 0.49$ for \cpd. The corresponding ISM neutral hydrogen column density amounts to $N_{\rm H}^{\rm ISM} = 2.8 \times 10^{21}$\,cm$^{-2}$. In the spectral fits, we thus requested a column density larger or equal to $N_{\rm H}^{\rm ISM}$. The best spectral fits are obtained for a two-temperature {\tt mekal} thermal plasma model \citep{MGvdO85, Ka92} with two independent absorbing columns. These fits indicate a soft (k$T\sim0.6$\,keV) slightly absorbed  plus a harder (k$T\sim1.0$\,keV) and more heavily absorbed component. However, they only provide an upper limit on the absorbing column associated to the soft component. As for \hda\ \citep{SSG04}, fixing this additional soft column to zero yields even better fits, characterized by more stable solutions.  The best-fit parameters are listed in Table \ref{tab: Xspec} and tend to indicate that the \cpd\ X-ray spectrum is significantly harder when the total flux is larger. More accurate information is however difficult to obtain since, as can be deduced from the modulations of the hardness ratios (Fig.~\ref{fig: hr}), the spectral variations are probably averaged out over the 30\,ks duration of a pointing. Unfortunately,  smaller bin sizes do not allow to obtain spectra of a sufficient quality to derive reliable constraints on the spectral properties.

\begin{figure}
\centering
\includegraphics[angle=-90,width=\columnwidth]{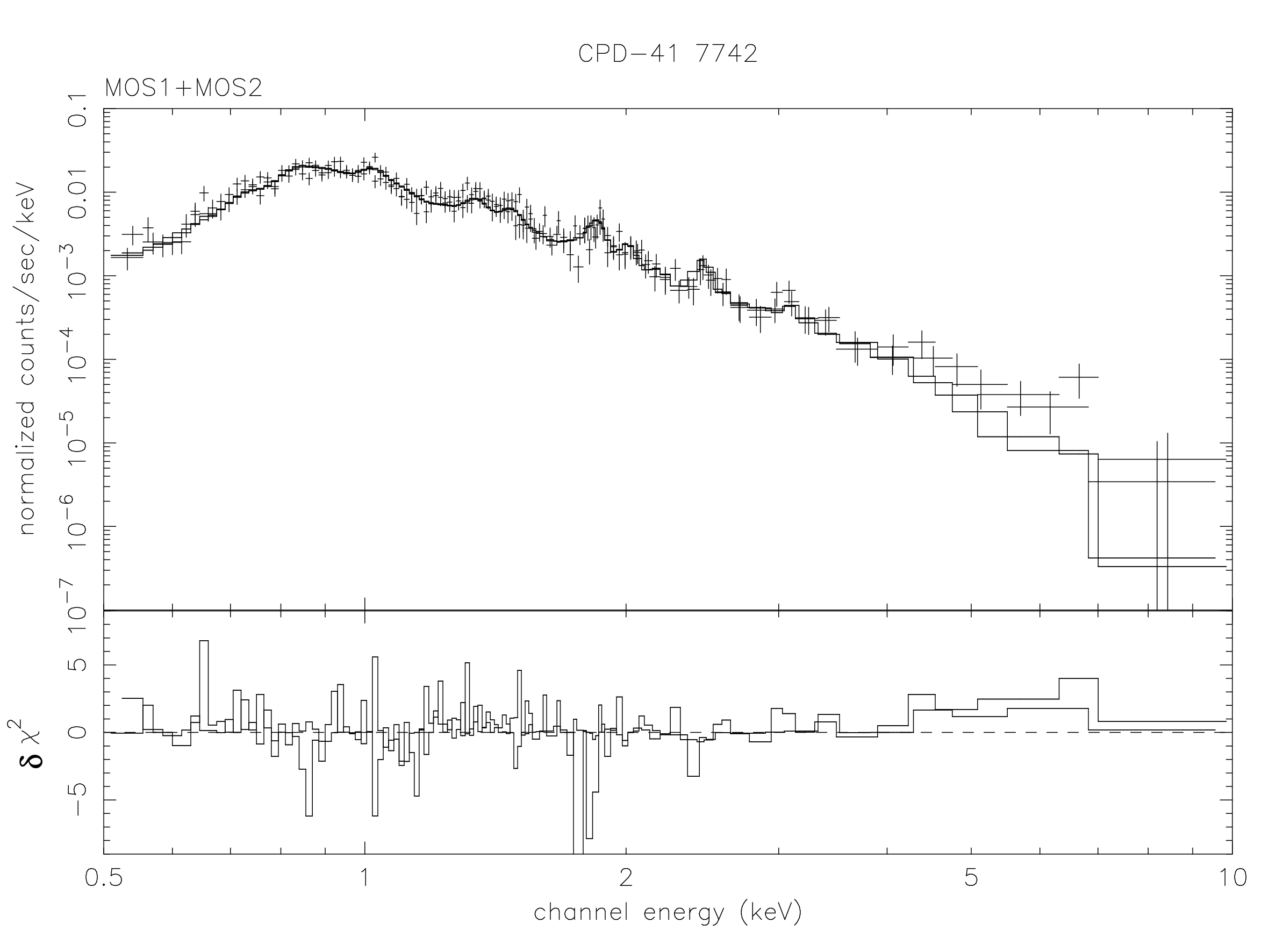}
\caption{ Least square simultaneous fits of the cumulated \mos1 and \mos2 spectra of \cpd\ with absorbed 2-T {\tt mekal} models. The bottom panel shows the contributions of individual bins to the $\chi^2$ of the fit. The contributions are carried over with the sign of the deviation (in the sense data minus model).\label{fig: Xspec_fig}}
\end{figure}

The combined \mos\ spectra obtained from the merging of the six \xmm\ observations are shown in Fig.~\ref{fig: Xspec_fig} together with the best fit 2-T model. Though the general quality of the fit is relatively good, the model tends to underestimate the fluxes at high energy ($>4$ \,keV). This could indicate the existence of a high energy component as well as the presence of the \ion{Fe}{K} line at 6.7\,keV. The merged spectra do unfortunately not have a sufficient quality at high energy to constrain this probable additional component.

\begin{table*}
\centering
\caption{Results of the simultaneous fits of the \mos1 and \mos2 spectra with {\sc xspec}. The model used is {\tt wabs$_\mathrm{ISM}$ * (wabs$_1$ * mekal$_1$ + wabs$_2$ * mekal$_2$)}. The term {\tt wabs$_\mathrm{ISM}$} was fixed to the interstellar value ($N_\mathrm{H, ISM}=0.28\times 10^{22}$\,cm$^{-2}$); {\tt wabs$_1$} was hold to zero ($N_\mathrm{H, 1}=0$\,cm$^{-2}$, see text). The first and second columns give the phase and the observation number. The next six columns (Cols. 3 to 8) provide the best-fit parameters while Col. 9 lists the corresponding reduced chi-square and the associated number of degrees of freedom (d.o.f). $N_\mathrm{H}$ yields the absorbing column (in units $10^{-22}$\,cm$^{-2}$), k$T$ is the model temperature (in keV) while $norm$ is the normalisation factor (expressed in $10^{-4}$\,cm$^{-5}$, $norm =\frac{10^{-14}}{4\pi d^2}\int n_\mathrm{e} n_\mathrm{H} dV$ with $d$, the distance to the source -- in cm --, $n_\mathrm{e}$ and $n_\mathrm{H}$, the electron and hydrogen number densities -- in cm$^{-3}$). Columns 10 to 13 provide the observed fluxes (in $10^{-14}$\,erg\,cm$^{-2}$\,s$^{-1}$) in the  0.5 - 10.0\,keV energy band and in the S$_\mathrm{X}$, M$_\mathrm{X}$ and H$_\mathrm{X}$ bands respectively. The last line of this table provides the best-fit parameters adjusted on the spectra extracted from the cumulated six pointings.} 
\label{tab: Xspec}
\begin{tabular}{c c c c c c c c c c c c c}
\hline
\hline
$\psi$ & Obs. \# & $N_\mathrm{H, 1}$ & k$T_1$ & $norm_1$ & $N_\mathrm{H, 2}$ & k$T_2$ & $norm_2$ & $\chi^2_{\nu}$ (d.o.f.) & $f_\mathrm{X}$ & $f_\mathrm{X,S}$ & $f_\mathrm{X,M}$  & $f_\mathrm{X,H}$ \\
$[1]$ & $[2]$ & $[3]$ & $[4]$ & $[5]$ & $[6]$ & $[7]$ & $[8]$ & $[9]$ & $[10]$ & $[11]$ & $[12]$ & $[13]$ \\
\hline
\vspace*{-3mm}  \\
0.113 &2 & 0.0                  & $0.62^{+.06}_{-.08}$ & $0.81^{+.15}_{-.13}$ & $0.79^{+.33}_{-.28}$ & $1.22^{+.26}_{-.17}$ & $2.71^{+.74}_{-.70}$  & 0.63 (59) & 21.0 & 5.6 & 11.5 & 3.8 \vspace*{1mm} \\
0.278 &5 & 0.0                  & $0.52^{+.14}_{-.18}$ & $0.65^{+.09}_{-.10}$ & $0.74^{+.25}_{-.19}$ & $0.97^{+.27}_{-.13}$ & $1.64^{+.42}_{-.52}$  & 1.21 (60) & 13.0 & 4.7 &  7.0 & 1.3 \vspace*{1mm} \\
0.468 &3 & 0.0                  & $0.52^{+.10}_{-.17}$ & $0.66^{+.09}_{-.10}$ & $0.76^{+.18}_{-.16}$ & $0.95^{+.12}_{-.13}$ & $2.28^{+.65}_{-.56}$  & 0.82 (81) & 15.4 & 4.9 &  8.7 & 1.8 \vspace*{1mm}\\ 
0.668 &6 & 0.0                  & $0.61^{+.05}_{-.14}$ & $0.75^{+.14}_{-.28}$ & $0.74^{+.41}_{-.36}$ & $0.93^{+.36}_{-.35}$ & $1.17^{+2.10}_{-.52}$ & 0.85 (64) & 12.7 & 5.3 &  6.5 & 0.9 \vspace*{1mm} \\
0.819 &1 & 0.0                  & $0.40^{+.10}_{-.08}$ & $0.76^{+.14}_{-.20}$ & $0.52^{+.27}_{-.21}$ & $0.81^{+.12}_{-.21}$ & $0.94^{+.48}_{-.20}$  & 1.08 (53) & 10.2 & 5.0 &  4.7 & 0.5 \vspace*{1mm} \\ 
0.930 &4 & 0.0                  & $0.35^{+.19}_{-.09}$ & $0.63^{+.30}_{-.23}$ & $0.46^{+.19}_{-.16}$ & $0.75^{+.14}_{-.08}$ & $1.61^{+.49}_{-.48}$  & 1.22 (46) & 11.8 & 5.2 &  6.0 & 0.6 \vspace*{1mm} \\ 
\hline 
\vspace*{-3mm}  \\
  \multicolumn{2}{c}{Merged} & 0.0                  & $0.59^{+.03}_{-.09}$ & $0.74^{+.06}_{-.08}$ & $0.73^{+.13}_{-.11}$ & $1.05^{+.11}_{-.08}$ & $1.39^{+.20}_{-.10}$  & 1.09 (209) & 14.0 & 5.2 &  7.3 & 1.5 \vspace*{1mm} \\ 
\hline 
\end{tabular}
\end{table*}

\section{\cpd\ X-ray properties \label{sect: Xmodel}}

\subsection{X-ray emission from the stellar components \label{sect: Xstars}}

The X-ray emission from massive stars presumably comes from shell collisions within the lower layers of their winds, which  result from the growing of radiatively-driven  wind instabilities \citep{FPP97}. It is expected that the bulk of the emission is produced in a zone extending to about five times the stellar radius. Within a binary system with an inclination close to 90\degr, we thus expect only a  small fraction of this extended emission zone to be occulted by the motion of one companion in front of the other. In consequence, because of the much larger emission zone, the eclipses in the X-ray domain are probably  not as clearly marked as in the optical.

However, the \cpd\ X-ray light curve (Fig.~\ref{fig: cpd42_lc5ks}) shows a clear decrease -- around $\psi=0.35$ -- almost perfectly synchronized with the optical secondary eclipse. 
This suggests a different geometry and, probably, the presence of a localized emission component, in addition to the intrinsic emission of the two stars. To match the observed light curve, this component should be occulted around $\psi=0.35$. It should thus  be associated either with the primary inner side, or with the secondary inner or outer sides. The emission level also appears to be lower between $\psi=0.6$ and 1.0, thus when the line of sight points both towards the primary inner side or the secondary outer side. The second possibility (i.e. an X-ray emission associated with the secondary inner side) therefore seems to best describe the main features of the X-ray light curve, at least qualitatively. In Sect. \ref{ssect: Xinteract}, we present a phenomenological model that  associates an extra X-ray emission  with the secondary inner side.

Using the relations of \citet{BSD97} and bolometric luminosities from  Table \ref{tab: Abs_Par}, we obtained X-ray luminosities of $\log(L_\mathrm{X})=31.51$ and $30.69$ (\ergs) respectively for the O9 and B1-1.5 components in the band $0.1-2.0$\,keV. Accounting for the distance modulus of the cluster $DM=11.07$, this corresponds to unabsorbed fluxes of $f_\mathrm{X}=9.99$ and $1.54\times 10^{-14}$\,\ergscm. Though the energy bands considered are slightly different, we can compare these predictions with the values obtained from the X-ray spectral fits (Table \ref{tab: Xflux}). It appears that, even at its minimum of emission ($\psi\sim0.82$), \cpd\ is at least twice brighter than expected from the \citeauthor{BSD97} relations.
Part of the gap between the observed and predicted values could however be filled by the following considerations. First, the dispersion around the \citeauthor{BSD97} relations is quite large and does not allow an accurate determination of the X-ray luminosities. Second, \citet{MSC84} reported that the winds from the main sequence B stars in \ngc\ are particularly strong. The B star in \cpd\ could thus have a particularly powerful wind for its spectral type, producing stronger shocks within its lower layers and, subsequently, an enhanced X-ray emission. \citet{SNG05} further reported that, in \ngc, the B stars seem to follow a brighter \lxlbol\ relation than predicted from \citet{BSD97}. From this new relation, the B1-1.5 component in \cpd\ could be at least three times brighter, yielding a luminosity of a few $10^{31}$\,\ergs.

\begin{table}
\centering
\caption{Unabsorbed fluxes (in $10^{-14}$\,erg\,cm$^{-2}$\,s$^{-1}$), i.e. fluxes corrected for the interstellar absorption ($N_\mathrm{H, ISM}=0.28\times 10^{22}$\,cm$^{-2}$), according to the best-fit models presented in Table \ref{tab: Xspec}. The last column gives the total X-ray luminosity (in \ergs) assuming a distance modulus $DM=11.07$.}
\label{tab: Xflux}
\begin{tabular}{c c c c c c c} 
\hline
\hline
$\psi$ & Obs. \# & $f_\mathrm{X}^\mathrm{unabs}$ & $f_\mathrm{X,S}^\mathrm{unabs}$& $f_\mathrm{X,M}^\mathrm{unabs}$ & $f_\mathrm{X,H}^\mathrm{unabs}$ &$\log(L_\mathrm{X})$\\
$[1]$ & $[2]$ & $[3]$ & $[4]$ & $[5]$ & $[6]$ & $[7]$ \\
\hline
 0.113 &2 & 36.2 & 16.4 & 15.8 & 4.0  & 32.06 \\
 0.278 &5 & 25.4 & 14.2 &  9.8 & 1.4  & 31.91 \\
 0.468 &3 & 28.7 & 14.8 & 12.1 & 1.8  & 31.96 \\ 
 0.668 &6 & 25.8 & 15.5 &  9.3 & 1.0  & 31.92 \\
 0.819 &1 & 23.8 & 16.4 &  6.9 & 0.5  & 31.88 \\ 
 0.930 &4 & 25.7 & 16.4 &  8.7 & 0.6  & 31.92 \\ 
\hline 
 \multicolumn{2}{c}{Merged} & 27.2 & 15.4 & 10.3 & 1.5  & 31.95 \\ 
\hline
\end{tabular}
\end{table}

\subsection{\cpd\ wind properties}
The wind properties of the two components of \cpd\ are not known. We however used the newly derived physical parameters of the stars to get an insight into their wind strengths. We estimated their mass-loss rates using the mass-loss recipes from \cite{VdKL00, VdKL01}. We obtained, for the primary, $\log(\dot{M}_1)=-7.06$ (\msol yr$^{-1}$). The temperature of the secondary component however falls within the bi-stability jump region. Using the recommendations from \citeauthor{VdKL00}, we estimated the position of the bi-stability jump to be located at about 22\,800\,K for the particular stellar parameters of the secondary. This puts the companion on the hot side of the jump, yielding thus $\log(\dot{M}_2)=-8.74$ (\msol yr$^{-1}$). We estimated the terminal wind velocities by first computing the escape velocities and then adopting the average ratio $v_\infty/v_\mathrm{esc}=2.6$ as appropriate for the winds of the stars on the hot side of the stability jump. We respectively obtained terminal velocities of $v_{\infty,1}=2380$\,\kms\ and $v_{\infty,2}=2150$\,\kms\  for the two components of \cpd. While these values are typical for O-type stars, the secondary terminal velocity seems quite large for a typical B1 dwarf. As stated above, \citet{MSC84} reported  particularly strong winds for the B dwarfs in \ngc. For example, they derived, for the single B1\,V star \object{CPD$-$41\degr\,7719}, a terminal wind velocity close to 2300\,\kms, thus very near our estimate for the B component in \cpd.

\subsection{A wind interaction in \cpd \label{ssect: Xinteract}}
Using the estimated wind parameters, we computed the position of the ram pressure equilibrium surface that typically indicates the location of a possible wind-wind collision. For this purpose, we adopted a $\beta=1$ velocity law, as appropriate for the hot star winds. Due to its larger mass-loss rate, the primary wind clearly overwhelms the secondary wind and no equilibrium is possible. In consequence, the O-star wind should crash into the B-star surface, preventing the secondary wind to develop towards the primary star. Under the above hypotheses, the primary wind luminosity at the distance of the secondary surface is about $\log(L_\mathrm{w,1})=\log(\frac{\dot{M}_1 v^2_1}{2})\sim 34.7$ (\ergs). Accounting for the secondary radius and its distance to the primary, a fraction of about 2.7\% of the O9\,V wind is intercepted by the secondary and we therefore expect the shocked plasma to be heated to temperatures of a few $10^7$\,K, thus generating a substantial amount of X-rays. According to the formalism of \citet{Uso92}, and using a primary wind pre-shock velocity of 1380\,\kms, the X-ray emission generated by such a wind-photosphere interaction should be about $\log(L_\mathrm{X})\approx 32.8$ (\ergs) for a purely radiative interaction (Usov's Eq. 80) and $\log(L_\mathrm{X})\approx 30.7$ (\ergs) in the adiabatic case (Usov's Eq. 79, adopting a solar chemical composition for the wind). Following \citet{SBP92}, the ratio between the characteristic cooling time and flow time is $\chi=t_\mathrm{cool}/t_\mathrm{flow}\approx 5.2$, indicating a mainly adiabatic collision.  

However, the interaction region is immersed in the intense UV photon field of the secondary. Inverse Compton cooling (Comptonization) could thus be significant, yielding a higher cooling rate, thus a lower value for the $\chi$ parameter. In addition, under the influence of the radiative pressure of the secondary, the acceleration of the primary wind may be slowed down (the so-called {\it radiative inhibition} effect, \citealt{StP94}). According to the formalism developed by these latter authors, the mass-loss rate on the axis should not be affected by more than 1\%. From a crude interpolation of their results, the primary wind velocity at the secondary surface might however be reduced by about one third. In consequence, the wind kinetical energy would be cut down by a factor of about two. Hence, radiative inhibition might significantly affect the value of the $\chi$ parameter, which depends on the fourth power of the velocity. Assuming a factor 2/3 on the velocity reached by the primary wind at the distance of the secondary surface, we obtain $\chi\approx1.0$. Both under the influence of Comptonization and of radiative inhibition, the shock region might thus shift towards the radiative regime.

Using the formalism of \citet{GOC97}, we also investigated the possibility to alter the wind-photosphere interaction by sudden radiative braking. In such a phenomenon, the wind of the primary star could be suddenly brought to a stop due to the radiative pressure coming from the companion. The main effect of radiative braking is to modify the position of the dynamical ram pressure equilibrium surface by pushing it further away from the secondary star. In certain cases, radiative braking could be strong enough to prevent the primary star wind to actually reach the secondary surface, thus yielding a wind-wind interaction structure rather than a wind-photosphere interaction. Adopting the known stellar parameters, we computed the radiative braking coefficient. We then used different values of the {\sc cak} parameters \citep{CAK75} appropriate for effective temperatures around 30\,kK. According to the values of these coefficients as given by different authors \citep{PPK86, SIH94, PSL00}, the radiative braking can, or can not, disrupt the wind-photosphere interaction. It is thus impossible to conclude on this point. However, even when the braking occurs, the interaction is moved only slightly away from the secondary star surface. Though the shock structure would be quite different, the geometry of the emitting region will probably remain rather similar, with an extra emission component mainly located close to the secondary inner surface.

\begin{figure*}
   \centering
   \includegraphics[angle=90,height=5cm]{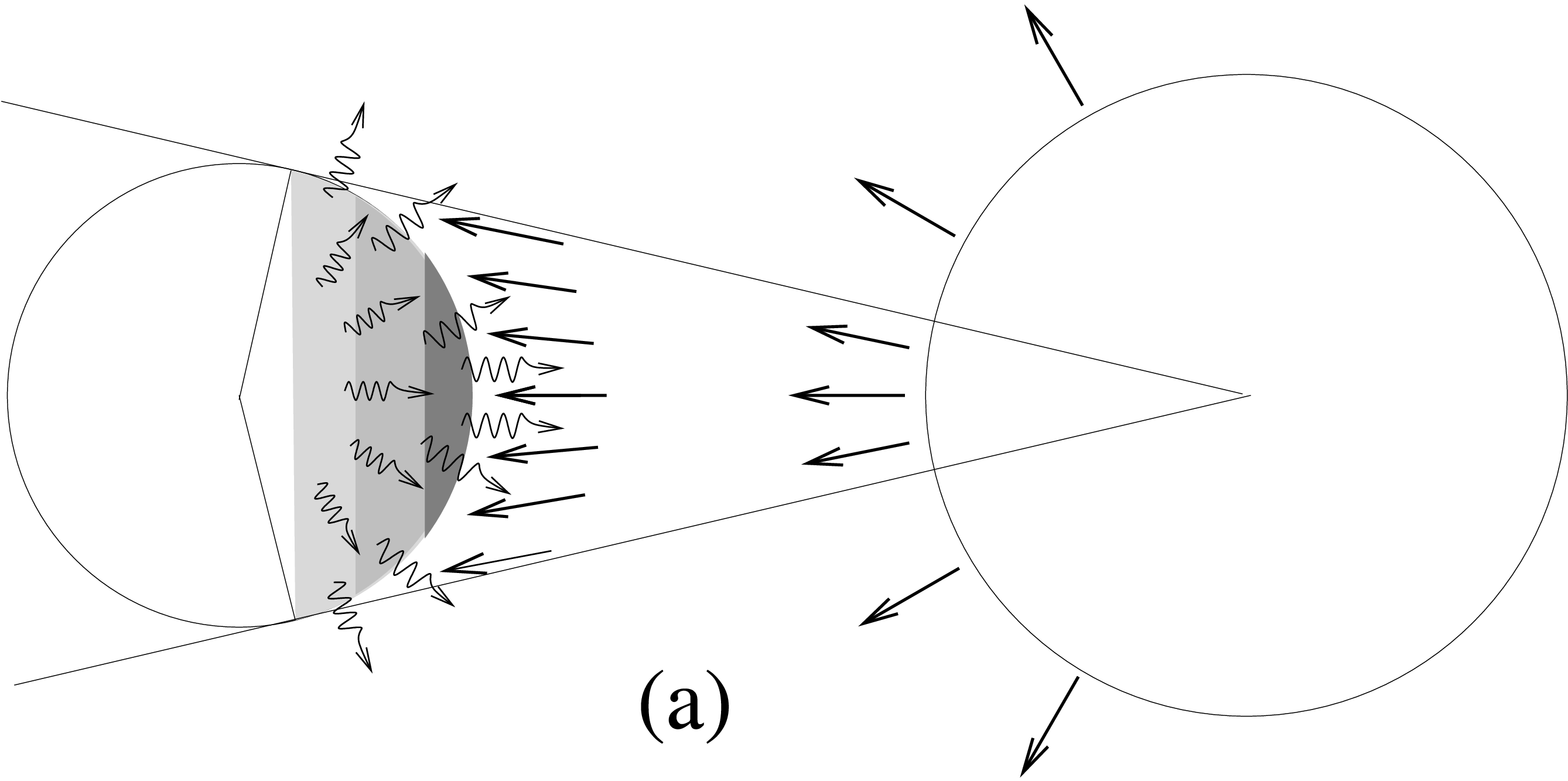}
\hspace*{1cm}
   \includegraphics[height=4.5cm]{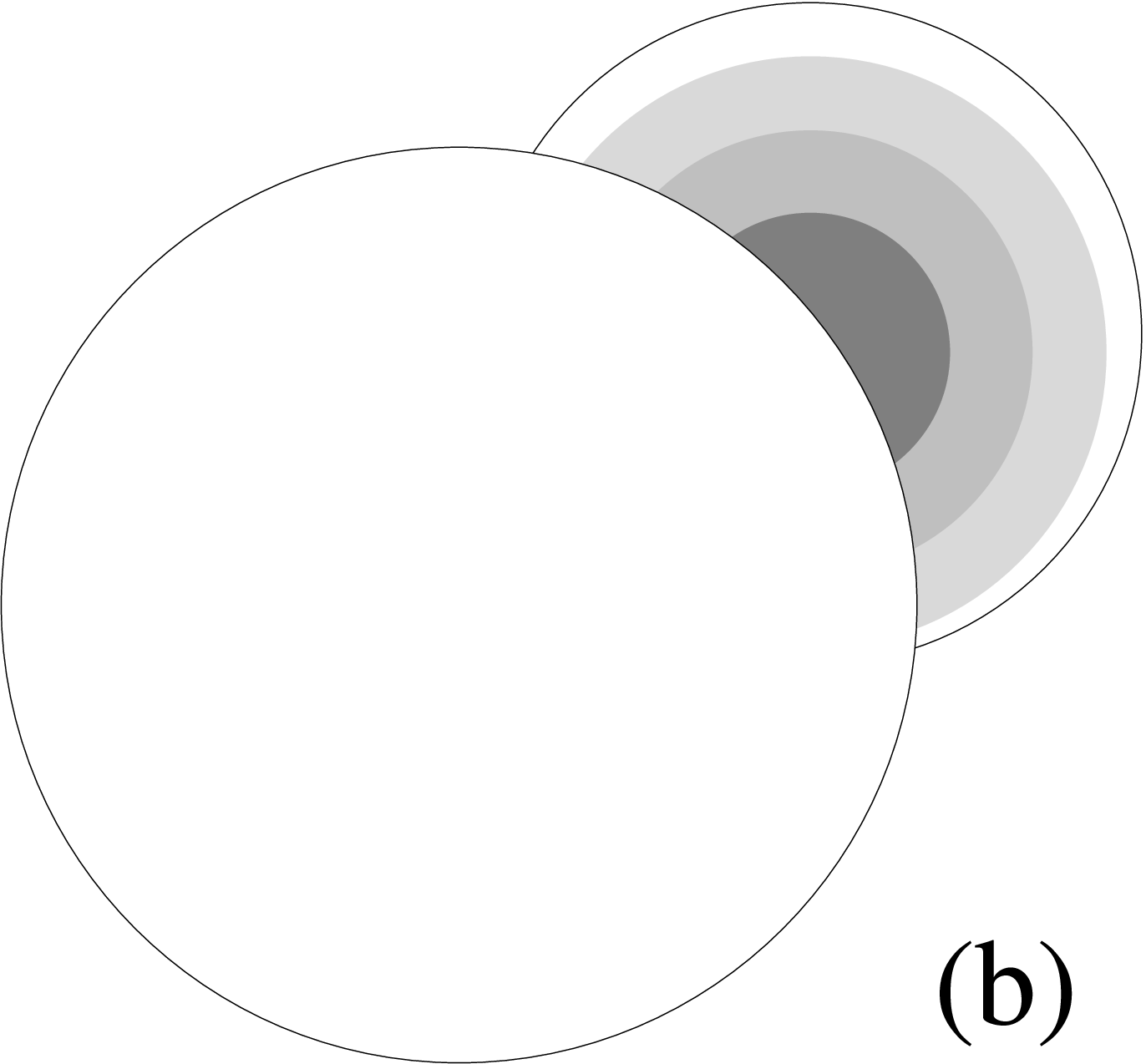}
   \caption{Schematic view of the geometrical wind-photosphere interaction model in \cpd. {\bf (a)} A view from above the orbital plane: the secondary star intercepts a small fraction of the primary wind, of which part of the kinetic energy is turned into heat.  {\bf (b)} A similar view along the line of sight. The different degrees of shading of the secondary surface represent the different X-ray luminosities, these latter being larger closer to the system axis.}
\label{fig: th_view}
\end{figure*}

\subsection*{A phenomenological model}

To estimate the influence of such a wind interaction on the observed X-ray light curve, we built a simple geometrical model presented in Fig.~\ref{fig: th_view}. We adopted a circular orbit, spherically symmetric stars and winds, and  a $\beta=1$ acceleration law for the primary wind. Assuming a totally radiative interaction, we considered that, when encountering the secondary star surface, the kinetic energy associated to the normal velocity component of the incident wind flow is totally dissipated into thermal energy. We thus computed the amount of energy re-emitted by each elements of the secondary surface. Then, accounting for the orbital inclination and the possible occultation by the primary star, we computed the  interaction contribution to the observed X-ray light curve. We noticed above the possibility of radiative braking to occur within the system. We caution however that it should not alter much this simple model. Indeed, in the case of a wind-wind collision, the interaction region should still be located near the secondary star surface, so that the geometry of the problem would be only slightly modified. The emission from the secondary shock would further be very limited. Indeed, so close to the surface, the radiative acceleration could not have been very efficient yet. The secondary wind velocity is thus probably of the order of the photospheric thermal velocity, therefore close to 20\,\kms. Under these hypotheses, the possible contribution of the secondary shock to the total X-ray emission would thus be about $10^{29}$\ergs, at least one order of magnitude below the other emission components.

\begin{figure}
   \centering
   \includegraphics[width=8cm]{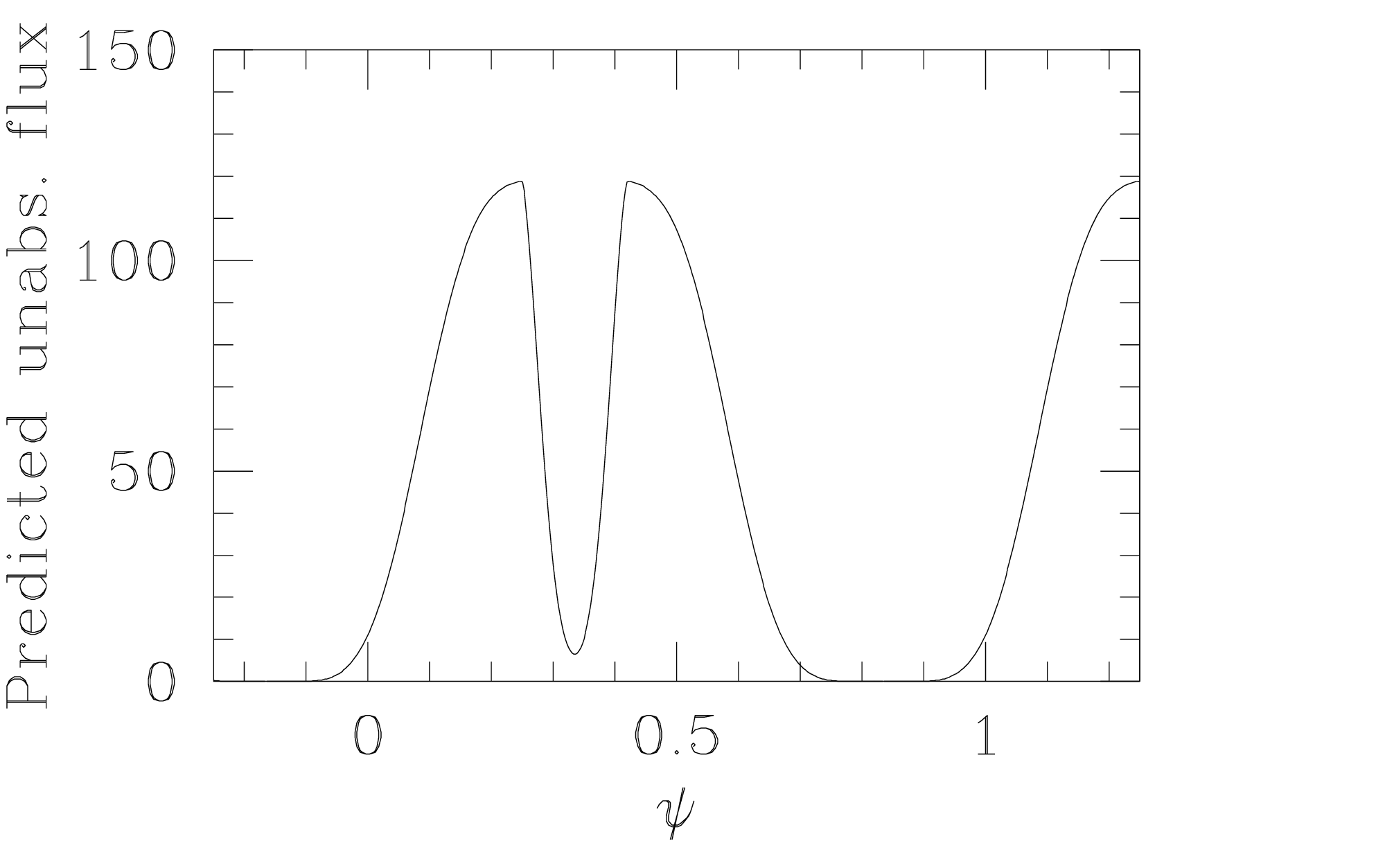}
   \includegraphics[width=8cm]{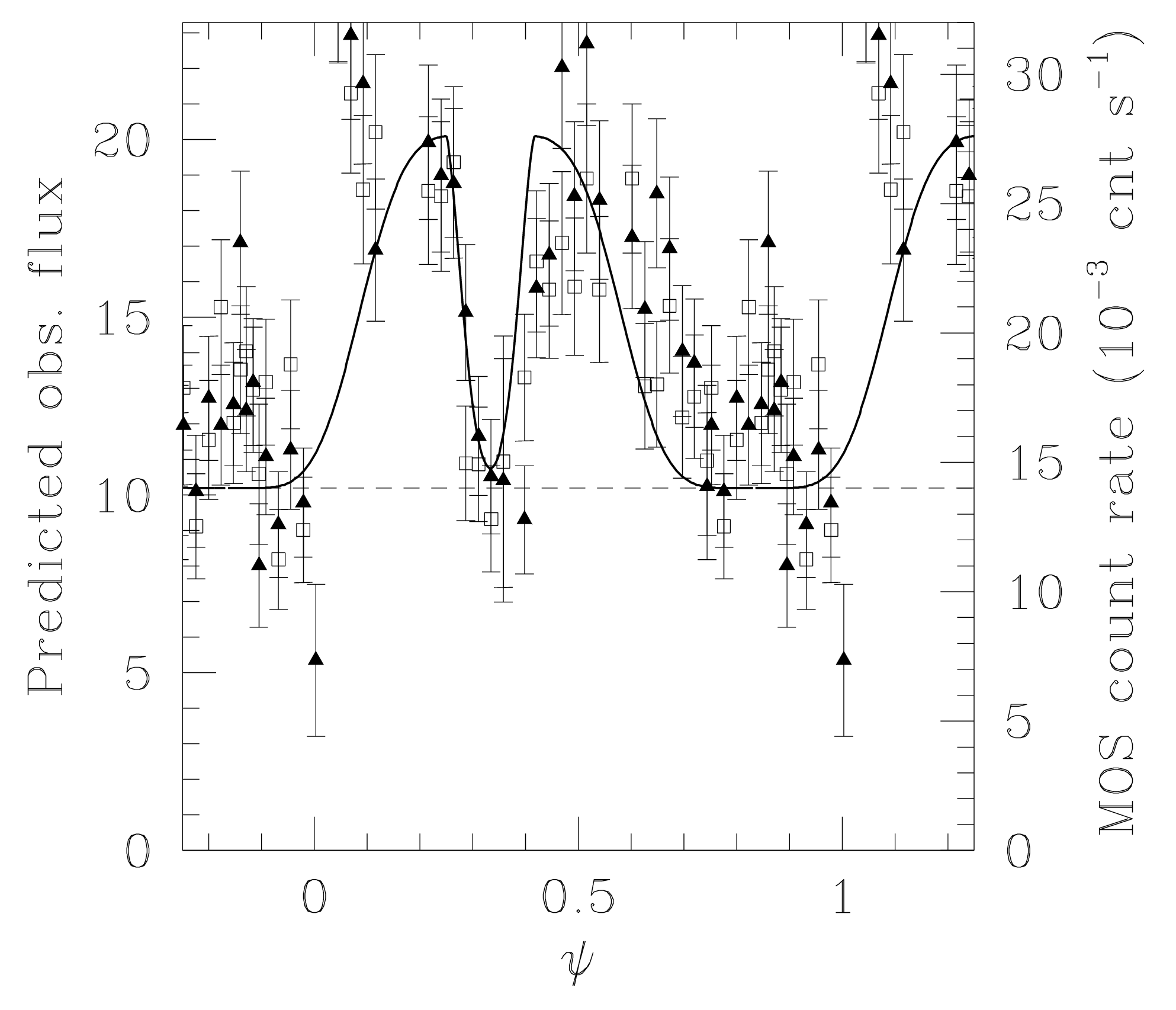}
   \caption{{\bf Upper panel:} Predicted unabsorbed flux emitted by a radiative wind-photosphere interaction in \cpd. {\bf Lower panel:} Tuned phenomenological model (thick line) overplotted on the observed X-ray light curves. Filled triangles and open squares respectively show the background-corrected  \epicmos1 and \mos2 count rates in the 0.5-10.0\,keV energy band. These count rates have been corrected for the limited size of the extraction region. Flux axes are in units of $10^{-14}$\,\ergscm\ in both panels. The dashed line gives the adopted intrinsic contribution from the two stellar components of \cpd. It acts as a pedestal.}
\label{fig: th_lc}
\end{figure}

The results of this simple model are  presented in Fig.~\ref{fig: th_lc} (upper panel) and provide an upper limit on the actual contribution of such an interaction. Indeed, as stated above, the interaction is probably not fully radiative, so that only a fraction of the incoming energy is effectively radiated.  Radiative inhibition might also reduce the wind velocity, giving rise to a weaker shock, hence to a weaker emission than considered here. In Fig.~\ref{fig: th_lc},  the occultation of the interaction zone by the primary is clearly seen ($\psi \sim 0.35$), while the interaction does not provide any contribution when the secondary is turning its outer side to the observer ($\psi\sim0.8-0.9$). In this simple form, this phenomenological model indeed  predicts  the higher emission slightly before and after the secondary eclipse, while the secondary inner side is facing the observer, and the lower emission state half a cycle later, when the interaction zone is hidden by the secondary body. It also reasonably reproduces the width of the observed {\it eclipse} in the X-ray light curve. 

In a second step, we try to provide a moderate tuning to the model, in order to investigate to which degree it can match the observed modulations. According to the model, no emission from the interaction is expected around $\psi=0.8-0.9$, and it should only provide a faint contribution at the time of the secondary minimum. At those particular phases, we thus probably observe the intrinsic emission of the two stars which, as explained above, is only slightly affected by the eclipses because of its wide extension.
Correcting the observed light curve for the limited encircled energy fraction, the intrinsic emission from the two stars gives about $14\times 10^{-3}$\,\cnts\ in the two \mos\ instruments. From Table \ref{tab: Xspec}, this approximately corresponds to an observed flux about $10.2\times 10^{-14}$\,\ergscm. We also note that the model provides unabsorbed fluxes while the observed count rates have suffered interstellar absorption. Comparing the values of the absorbed and unabsorbed fluxes (Tables \ref{tab: Xspec} and \ref{tab: Xflux}), we estimate the ISM material to absorb about half of the flux at the considered energy.  

At this stage, the model predicts an emission rate still much larger than the observed emission. To properly match the observations, we have to divide the predicted flux again by a factor of 6, so that the maximum contribution of the interaction zone to the observed flux is now about $10\times 10^{-14}$\,\ergscm. This last step finds a relative justification in the fact that, as discussed above, the present purely radiative model only provides an upper limit to the X-ray emission emerging from the interaction zone.  In addition, effects that might reduce the shock strength, such as radiative inhibition, are not accounted for. Assuming radiative braking to take place, one might also think that some emission may originate from the trailing arm of the shock cone. In such a configuration, the extra emission from the collision would not drop to zero around $\psi\sim0.8-0.9$. 
In the current fully radiative model, the plasma immediately cools down after the shock. By nature, it could thus not produce an extra-emission at these particular phases. Such a contribution from the arms of the shock cone would however be less affected by the eclipses in the system and, as a first approximation,  one can consider that it has been accounted for in the empirical pedestal adopted in Fig.~\ref{fig: th_lc}. This latter is indeed higher than expected from the sole \citet{BSD97} relations for the intrinsic emission of the early-type stars in the system. 

It is clear that the reality is probably different from this idealized situation.  It is not our purpose to over-interpret the present model; our aim was to show that, using reasonable assumptions, an interaction region located on, or near, the secondary surface can reproduce the main features of the observed X-ray light curve.

From Fig.~\ref{fig: th_lc}, the time of the beginning of the X-ray eclipse is well reproduced by our model. The right wing of the eclipse is however slightly larger, suggesting that the interaction is more extended on the surface side opposite to the orbital motion.  Similarly, the drop in emission around $\psi\approx 0.6$ occurs slightly later than expected from our model, which means that the X-ray emitting region should remain visible slightly longer, a condition which does not tally with the previous suggestion. Clearly, Fig.~\ref{fig: th_lc} shows that  the observed modulations in the X-ray light curves is dominated by an extra-emission component associated with the secondary inner surface. However, the details of the phenomenon could be more complicated as suggested by the observed delays  in the rising and falling branches near $\psi=0.4$ and 0.6.
Finally, the hardness ratio curves indicate that the hardest emission is observed at the time of the two emission maxima. This is expected if the extra emission is produced in a wind interaction region, which provides typically harder X-rays than the intrinsic emission from the stars.


\section{Final remarks and conclusions  \label{sect: ccl}}
In the first part of this paper, we presented optical photometry of \cpd. Adopting the period obtained from \citetalias{SHRG03}, the analysis of the system light curves indicates that \cpd\ is a well detached system with an inclination close to 77\degr. The obtained curves display two eclipses with a separation slightly different from half an orbital cycle, thus indicating a small eccentricity, in agreement with the results of \citetalias{SHRG03}. Combining the spectroscopic and photometric analyses, we derived the absolute physical parameters of the stellar components and confirmed that the system is formed by two dwarf early-type stars with masses, sizes and luminosities  relatively close to typical values expected both from observational and theoretical works.

The photometric and spectroscopic data sets however provide discrepant values for the longitude of periastron. Independent observations by \citet{StB04} also tend to indicate either a periastron argument close to 90 or 270\degr\ or a zero eccentricity. Their light curves, obtained over at least two years, also display intriguing signs of variability. 
Clearly, further observations are needed to elucidate these apparent discrepancies.

In the second part of the present paper, we focused on recent  \xmm\ X-ray observations of the system. The X-ray emission from \cpd\ is well described by a two-temperature thermal plasma model with energies close to 0.6 and 1.0\,keV, thus slightly harder than typical emission from early-type stars. The X-ray light curve of the system is clearly variable both in the total  and in the different energy ranges; the emission level is higher when the primary is in front of the secondary. During the high state, the system shows a drop of its X-ray emission that almost exactly matches the optical secondary eclipse. Assuming that the X-ray light curve is reproductible, we interpreted this as the signature of a wind interaction phenomenon in which the overwhelming primary wind crashes into the secondary star surface. Alternatively, the wind-photosphere interaction could be altered by sudden radiative braking, yielding a wind-wind interaction located close to the secondary surface, and displaying thus a similar geometry. We expect this phenomenon  to produce a substantial amount of X-rays, which could be the major source for the observed modulations in the \epicmos\ light curves. As a next step, we built a simple phenomenological model that associates an extra X-ray emission component with the inner side of the secondary star surface. Though limited by some simplifying assumptions, this model renders the main properties of the observed variations  and lends thus further support to our interpretation of the X-ray light curve. 

At this stage, several important questions remain however unanswered. The exact influence of the wind interaction, and of the generated  X-ray emission, on the secondary surface properties is very difficult to estimate. We carefully inspected the high resolution high signal to noise spectra from \citetalias{SHRG03} but could not find any systematic differences in the secondary spectra obtained when this star is showing either its inner or its outer face to the observer. 
As a final check, we put a point-like X-ray source at a distance of $1.1\times R_2$ from the center of the secondary star on the system axis. We assigned to this source a luminosity of $10^{33}$\,\ergs, which is probably typical of the wind interaction taking place in the system. The additional heating of the secondary star surface elements closest to the X-ray source amounts to a few tens of Kelvin. For comparison, the heating of the same surface elements by the radiation of the primary component, is about 2000-2500\,K. 
This clearly suggests that the heating of the secondary surface by the nearby interaction should be limited. 

Formed by an O9 plus a B1-1.5 dwarf, \cpd\ {\it a priori} seemed to be an ordinary, well detached system. We however showed that it probably harbours a wind-wind or wind-photosphere interaction. Such a phenomenon could be quite common among close early-type systems. It is thus of a particular importance to evaluate its possible impact on the determination of the physical parameters obtained using different observational methods. 
The possible variable activity of \cpd\ is an additional motivation to accumulate more data on this particularly interesting early-type binary system. 

Finally, the present set of observations provides X-ray light curves that cover almost the full orbital cycle of \cpd\ with reasonable signal-to-noise and time resolution.  As discussed in Sect. \ref{ssect: Xinteract}, different physical phenomena (radiative inhibition, radiative braking, Comptonization, ...) probably affect the shock structure and, hence, the exact amount of X-ray emission generated by the wind interaction.
  The development of appropriate tools, both theoretical and numerical, to analyse such high quality X-ray light curves is probably one of the challenges that the new generation of X-ray stellar scientists will have to face in the coming decade, especially to prepare the ground for the next generation of large X-ray observatories.

\begin{acknowledgements}
It is a pleasure to thank Dr. I.I. Antokhin for fruitful discussions.
The one month run at the Bochum telescope has been made
possible thanks to a `cr\'edit aux chercheurs' 
from the Belgian FNRS. Our Bochum negotiators,
H.G. Grothues and R.J. Dettmar, are warmly thanked
for their open-minded efficiency.
We also acknowledge support from the PRODEX XMM-OM and Integral Projects, as well as contracts P4/05 and P5/36 `P\^ole d'Attraction Interuniversitaire' (Belgium).
EA acknowledges support from the Russian Foundation for Basic
Research (project No 02-02-17524) and the Russian LSS (project
No 388.2003.2). 
\end{acknowledgements}

\bibliographystyle{aa}

\bibliography{/datas6/XMM_CAT_PAPER/ngc6231_Xcat}

\begin{thebibliography}{66}
\expandafter\ifx\csname natexlab\endcsname\relax\def\natexlab#1{#1}\fi

\bibitem[{{Antokhina}(1988)}]{Ant88}
{Antokhina}, E.~A. 1988, \azh, 65, 1164

\bibitem[{{Antokhina}(1996)}]{Ant96}
{Antokhina}, E.~A. 1996, Astronomy Reports, 40, 483

\bibitem[{{Arnaud}(1996)}]{arn96}
{Arnaud}, K.~A. 1996, in ASP Conf. Ser., Vol. 101, Astronomical Data Analysis
  Software and Systems V, ed. G.~{Jacoby} \& J.~{Barnes}, 17

\bibitem[{{Balona} \& {Laney}(1995)}]{BL95}
{Balona}, L.~A., \& {Laney}, C.~D. 1995, \mnras, 276, 627

\bibitem[{{Barr}(1908)}]{Barr08}
{Barr}, J. 1908, \jrasc, 2, 70

\bibitem[{{Batten}(1983)}]{Bat83}
{Batten}, A.~H. 1983, \jrasc, 77, 95

\bibitem[{{Batten}(1988)}]{Bat88}
{Batten}, A.~H. 1988, \pasp, 100, 160

\bibitem[{{Baume} {et~al.}(1999){Baume}, {V{\' a}zquez}, \&
  {Feinstein}}]{BVF99}
{Baume}, G., {V{\' a}zquez}, R.~A., \& {Feinstein}, A. 1999, \aaps, 137, 233

\bibitem[{{Bell} {et~al.}(1987){Bell}, {Hilditch}, \& {Adamson}}]{BHA87}
{Bell}, S.~A., {Hilditch}, R.~W., \& {Adamson}, A.~J. 1987, \mnras, 225, 961

\bibitem[{{Bergh\"ofer} \& {Schmitt}(1994)}]{BS94}
{Bergh\"ofer}, T.~W., \& {Schmitt}, J.~H.~M.~M. 1994, \apss, 221, 309

\bibitem[{{Bergh\"ofer} {et~al.}(1997){Bergh\"ofer}, {Schmitt}, {Danner}, \&
  {Cassinelli}}]{BSD97}
{Bergh\"ofer}, T.~W., {Schmitt}, J.~H.~M.~M., {Danner}, R., \& {Cassinelli},
  J.~P. 1997, \aap, 322, 167

\bibitem[{{Bianchi} \& {Garcia}(2002)}]{BG02}
{Bianchi}, L., \& {Garcia}, M. 2002, \apj, 581, 610

\bibitem[{{Burkholder} {et~al.}(1997){Burkholder}, {Massey}, \&
  {Morrell}}]{BMM97}
{Burkholder}, V., {Massey}, P., \& {Morrell}, N. 1997, \apj, 490, 328

\bibitem[{{Castor} {et~al.}(1975){Castor}, {Abbott}, \& {Klein}}]{CAK75}
{Castor}, J.~I., {Abbott}, D.~C., \& {Klein}, R.~I. 1975, \apj, 195, 157

\bibitem[{{Crowther} {et~al.}(2002){Crowther}, {Hillier}, {Evans}, {Fullerton},
  {De Marco}, \& {Willis}}]{CHE02}
{Crowther}, P.~A., {Hillier}, D.~J., {Evans}, C.~J., {et~al.} 2002, \apj, 579,
  774

\bibitem[{{Diaz-Cordoves} \& {Gimenez}(1992)}]{DCG92}
{Diaz-Cordoves}, J., \& {Gimenez}, A. 1992, \aap, 259, 227

\bibitem[{{Diaz-Cordoves} {et~al.}(1995){Diaz-Cordoves}, {Claret}, \&
  {Gimenez}}]{DCCG95}
{Diaz-Cordoves}, J., {Claret}, A., \& {Gimenez}, A. 1995, \aaps, 110, 329

\bibitem[{{Feldmeier} {et~al.}(1997){Feldmeier}, {Puls}, \&
  {Pauldrach}}]{FPP97}
{Feldmeier}, A., {Puls}, J., \& {Pauldrach}, A.~W.~A. 1997, \aap, 322, 878

\bibitem[{{Fracastoro}(1979)}]{Fra79}
{Fracastoro}, M.~G. 1979, \aap, 78, 112

\bibitem[{{Gayley} {et~al.}(1997){Gayley}, {Owocki}, \& {Cranmer}}]{GOC97}
{Gayley}, K.~G., {Owocki}, S.~P., \& {Cranmer}, S.~R. 1997, \apj, 475, 786

\bibitem[{{Gies}(2003)}]{Gie03}
{Gies}, D.~R. 2003, in IAU Symposium, Vol. 212, A Massive Star Odyssey: from
  main sequence to supernova, ed. K.~{van der Hucht}, A.~{Herrero}, \&
  C.~{Esteban}, 91

\bibitem[{{Herrero}(2003)}]{He03}
{Herrero}, A. 2003, in IAU Symposium, Vol. 212, A Massive Star Odyssey: from
  main sequence to supernova, ed. K.~{van der Hucht}, A.~{Herrero}, \&
  C.~{Esteban}, 3

\bibitem[{{Herrero} {et~al.}(1992){Herrero}, {Kudritzki}, {Vilchez}, {Kunze},
  {Butler}, \& {Haser}}]{HKV92}
{Herrero}, A., {Kudritzki}, R.~P., {Vilchez}, J.~M., {et~al.} 1992, \aap, 261,
  209

\bibitem[{{Herrero} {et~al.}(2002){Herrero}, {Puls}, \& {Najarro}}]{HPN02}
{Herrero}, A., {Puls}, J., \& {Najarro}, F. 2002, \aap, 396, 949

\bibitem[{{Hill} \& {Holmgren}(1995)}]{HiH95}
{Hill}, G., \& {Holmgren}, D.~E. 1995, \aap, 297, 127

\bibitem[{{Himmelblau}(1971)}]{Him71}
{Himmelblau}, D.~M. 1971, Applied Nonlinear Programming (New-York:
  McGraw--Hill)

\bibitem[{{Howarth}(1993)}]{How93}
{Howarth}, I.~D. 1993, The Observatory, 113, 75

\bibitem[{{Howarth} \& {Prinja}(1989)}]{HP89}
{Howarth}, I.~D., \& {Prinja}, R.~K. 1989, \apjs, 69, 527

\bibitem[{{Humphreys} \& {McElroy}(1984)}]{HM84}
{Humphreys}, R.~M., \& {McElroy}, D.~B. 1984, \apj, 284, 565

\bibitem[{{Jansen} {et~al.}(2001){Jansen}, {Lumb}, {Altieri}, {Clavel}, {Ehle},
  {Erd}, {Gabriel}, {Guainazzi}, {Gondoin}, {Much}, {Munoz}, {Santos},
  {Schartel}, {Texier}, \& {Vacanti}}]{Jansen01_xmm}
{Jansen}, F., {Lumb}, D., {Altieri}, B., {et~al.} 2001, \aap, 365, L1

\bibitem[{{Kaastra}(1992)}]{Ka92}
{Kaastra}, J. 1992, {An X-Ray Spectral Code for Optically Thin Plasmas},
  (Internal SRON-Leiden Report, updated version 2.0)

\bibitem[{{Kallrath} \& {Linnell}(1987)}]{KL87}
{Kallrath}, J., \& {Linnell}, A.~P. 1987, \apj, 313, 346

\bibitem[{{Manfroid}(1993)}]{Man93}
{Manfroid}, J. 1993, \aap, 271, 714

\bibitem[{{Manfroid}(1995)}]{Man95}
{Manfroid}, J. 1995, \aaps, 113, 587

\bibitem[{{Manfroid} {et~al.}(2001){Manfroid}, {Royer}, {Rauw}, \&
  {Gosset}}]{MRR01}
{Manfroid}, J., {Royer}, P., {Rauw}, G., \& {Gosset}, E. 2001, in ASP Conf.
  Ser., Vol. 238, Astronomical Data Analysis Software and Systems X, ed.
  F.~{Harnden}, F.~{Primini}, \& H.~{Payne}, 373

\bibitem[{{Martins} {et~al.}(2002){Martins}, {Schaerer}, \& {Hillier}}]{MSH02}
{Martins}, F., {Schaerer}, D., \& {Hillier}, D.~J. 2002, \aap, 382, 999

\bibitem[{{Massa} {et~al.}(1984){Massa}, {Savage}, \& {Cassinelli}}]{MSC84}
{Massa}, D., {Savage}, B.~D., \& {Cassinelli}, J.~P. 1984, \apj, 287, 814

\bibitem[{{Mewe} {et~al.}(1985){Mewe}, {Gronenschild}, \& {van den
  Oord}}]{MGvdO85}
{Mewe}, R., {Gronenschild}, E.~H.~B.~M., \& {van den Oord}, G.~H.~J. 1985,
  \aaps, 62, 197

\bibitem[{{Pauldrach} {et~al.}(1986){Pauldrach}, {Puls}, \&
  {Kudritzki}}]{PPK86}
{Pauldrach}, A., {Puls}, J., \& {Kudritzki}, R.~P. 1986, \aap, 164, 86

\bibitem[{{Popper} \& {Hill}(1991)}]{PoH91}
{Popper}, D.~M., \& {Hill}, G. 1991, \aj, 101, 600

\bibitem[{{Puls} {et~al.}(2000){Puls}, {Springmann}, \& {Lennon}}]{PSL00}
{Puls}, J., {Springmann}, U., \& {Lennon}, M. 2000, \aaps, 141, 23

\bibitem[{{Rauw} {et~al.}(2001){Rauw}, {Sana}, {Antokhin}, {Morrell},
  {Niemela}, {Albacete Colombo}, {Gosset}, \& {Vreux}}]{RSA01}
{Rauw}, G., {Sana}, H., {Antokhin}, I.~I., {et~al.} 2001, \mnras, 326, 1149

\bibitem[{{Royer} {et~al.}(1998){Royer}, {Vreux}, \& {Manfroid}}]{RVM98}
{Royer}, P., {Vreux}, J.-M., \& {Manfroid}, J. 1998, \aaps, 130, 407

\bibitem[{{Sana} {et~al.}(2003){Sana}, {Hensberge}, {Rauw}, \&
  {Gosset}}]{SHRG03}
{Sana}, H., {Hensberge}, H., {Rauw}, G., \& {Gosset}, E. 2003, \aap, 405, 1063

\bibitem[{{Sana} {et~al.}(2004){Sana}, {Stevens}, {Gosset}, {Rauw}, \&
  {Vreux}}]{SSG04}
{Sana}, H., {Stevens}, I.~R., {Gosset}, E., {Rauw}, G., \& {Vreux}, J.-M. 2004,
  \mnras, 350, 809

\bibitem[{{Sana} {et~al.}(2005a){Sana}, {Gosset}, {Rauw}, {Sung}, \&
  {Vreux}}]{SGR05}
{Sana}, H., {Gosset}, E., {Rauw}, G., {Sung}, H., \& {Vreux}, J.-M. 2005a, 
 \aap, in press

\bibitem[{{Sana} {et~al.}(2005b){Sana}, {Naz\'e}, {Gosset}, {Rauw}, {Sung}, \&
  {Vreux}}]{SNG05}
{Sana}, H., {Naz\'e}, Y., {Gosset}, E., {et~al.} 2005b, in {Massive Stars in
  Interacting Binaries}, ed. A.~{Moffat} \& N.~{St-Louis}, ASP Conf. Ser., 5p.,
  in press

\bibitem[{{Schaller} {et~al.}(1992){Schaller}, {Schaerer}, {Meynet}, \&
  {Maeder}}]{SSM92}
{Schaller}, G., {Schaerer}, D., {Meynet}, G., \& {Maeder}, A. 1992, \aaps, 96,
  269

\bibitem[{{Schmidt-Kaler}(1982)}]{SK82}
{Schmidt-Kaler}, T. 1982, {Landolt-B\"ornstein, Numerical Data and Functional
  Relationships in Science and Technology, New Series, Group VI}, Vol.~2b,
  {Physical Parameters of the Stars} (Berlin: Springer-Verlag)

\bibitem[{{Shimada} {et~al.}(1994){Shimada}, {Ito}, {Hirata}, \&
  {Horaguchi}}]{SIH94}
{Shimada}, M.~R., {Ito}, M., {Hirata}, B., \& {Horaguchi}, T. 1994, in IAU
  Symp., Vol. 162, Pulsation; Rotation; and Mass Loss in Early-Type Stars, ed.
  L.~{Balona}, H.~{Henrichs}, \& J.~{Contel}, 487

\bibitem[{{Simon} {et~al.}(1994){Simon}, {Sturm}, \& {Fiedler}}]{SSF94}
{Simon}, K.~P., {Sturm}, E., \& {Fiedler}, A. 1994, \aap, 292, 507

\bibitem[{{Sterken} \& {Bouzid}(2004)}]{StB04}
{Sterken}, C., \& {Bouzid}, M.~Y. 2004, in Rev. Mex. Astron. Astrofis., Conf.
  Ser., Vol.~20, Compact Binaries in the Galaxy and Beyond, ed. G.~{Tovmassian}
  \& E.~{Sion}, 79

\bibitem[{{Stetson}(1987)}]{Ste87}
{Stetson}, P.~B. 1987, \pasp, 99, 191

\bibitem[{{Stevens} \& {Pollock}(1994)}]{StP94}
{Stevens}, I.~R., \& {Pollock}, A.~M.~T. 1994, \mnras, 269, 226

\bibitem[{{Stevens} {et~al.}(1992){Stevens}, {Blondin}, \& {Pollock}}]{SBP92}
{Stevens}, I.~R., {Blondin}, J.~M., \& {Pollock}, A.~M.~T. 1992, \apj, 386, 265

\bibitem[{{Struve}(1948)}]{Str48}
{Struve}, O. 1948, \pasp, 60, 160

\bibitem[{{Sung} {et~al.}(1998){Sung}, {Bessell}, \& {Lee}}]{SBL98}
{Sung}, H., {Bessell}, M.~S., \& {Lee}, S. 1998, \aj, 115, 734

\bibitem[{{Turner} {et~al.}(2001){Turner}, {Abbey}, {Arnaud}, {Balasini},
  {Barbera}, {Belsole}, {Bennie}, {Bernard}, {Bignami}, {Boer}, {Briel},
  {Butler}, {Cara}, {Chabaud}, {Cole}, {Collura}, {Conte}, {Cros}, {Denby},
  {Dhez}, {Di Coco}, {Dowson}, {Ferrando}, {Ghizzardi}, {Gianotti}, {Goodall},
  {Gretton}, {Griffiths}, {Hainaut}, {Hochedez}, {Holland}, {Jourdain},
  {Kendziorra}, {Lagostina}, {Laine}, {La Palombara}, {Lortholary}, {Lumb},
  {Marty}, {Molendi}, {Pigot}, {Poindron}, {Pounds}, {Reeves}, {Reppin},
  {Rothenflug}, {Salvetat}, {Sauvageot}, {Schmitt}, {Sembay}, {Short},
  {Spragg}, {Stephen}, {Str{\" u}der}, {Tiengo}, {Trifoglio}, {Tr{\" u}mper},
  {Vercellone}, {Vigroux}, {Villa}, {Ward}, {Whitehead}, \&
  {Zonca}}]{Turner01_mos}
{Turner}, M.~J.~L., {Abbey}, A., {Arnaud}, M., {et~al.} 2001, \aap, 365, L27

\bibitem[{{Usov}(1992)}]{Uso92}
{Usov}, V.~V. 1992, \apj, 389, 635

\bibitem[{{Vacca} {et~al.}(1996){Vacca}, {Garmany}, \& {Shull}}]{VGS96}
{Vacca}, W.~D., {Garmany}, C.~D., \& {Shull}, J.~M. 1996, \apj, 460, 914

\bibitem[{{van Hamme}(1993)}]{vHa93}
{van Hamme}, W. 1993, \aj, 106, 2096

\bibitem[{{Vaz} {et~al.}(1997){Vaz}, {Cunha}, {Vieira}, \& {Myrrha}}]{VCV97}
{Vaz}, L.~P.~R., {Cunha}, N.~C.~S., {Vieira}, E.~F., \& {Myrrha}, M.~L.~M.
  1997, \aap, 327, 1094

\bibitem[{{Vink} {et~al.}(2000){Vink}, {de Koter}, \& {Lamers}}]{VdKL00}
{Vink}, J.~S., {de Koter}, A., \& {Lamers}, H.~J.~G.~L.~M. 2000, \aap, 362, 295

\bibitem[{{Vink} {et~al.}(2001){Vink}, {de Koter}, \& {Lamers}}]{VdKL01}
{Vink}, J.~S., {de Koter}, A., \& {Lamers}, H.~J.~G.~L.~M. 2001, \aap, 369, 574

\bibitem[{{Wilson}(1979)}]{Wil79}
{Wilson}, R.~E. 1979, \apj, 234, 1054

\bibitem[{{Zinnecker}(2003)}]{Zin03}
{Zinnecker}, H. 2003, in IAU Symposium, Vol. 212, A Massive Star Odyssey: from
  main sequence to supernova, ed. K.~{van der Hucht}, A.~{Herrero}, \&
  C.~{Esteban}, 80

\end{thebibliography}
\end{document}